\documentclass[sigconf]{aamas}  % do not change this line!

\usepackage{booktabs}
\usepackage{subfigure}
\usepackage{algorithm}
\usepackage{algorithmicx}
\usepackage{algpseudocode}
\usepackage{amsmath}
\usepackage{subfigure}

\usepackage{booktabs}
\usepackage{colortbl}
\definecolor{mygray}{gray}{.9}

\setcopyright{ifaamas}  % do not change this line!
\acmDOI{doi}  % do not change this line!
\acmISBN{}  % do not change this line!
\acmConference[AAMAS'18]{Proc.\@ of the 17th International Conference on Autonomous Agents and Multiagent Systems (AAMAS 2018), M.~Dastani, G.~Sukthankar, E.~Andre, S.~Koenig (eds.)}{July 2018}{Stockholm, Sweden}  % do not change this line!
\acmYear{2018}  % do not change this line!
\copyrightyear{2018}  % do not change this line!
\acmPrice{}  % do not change this line!

\begin{document}

\title{An Optimal Rewiring Strategy for Reinforcement Social Learning in Cooperative Multiagent Systems}  % put your title here!
%\titlenote{Produces the permission block, and copyright information}

% AAMAS: as appropriate, uncomment one subtitle line; check the CFP
%\subtitle{Extended Abstract}
%\subtitle{Industrial Applications Track}
%\subtitle{Socially Interactive Agents Track}
%\subtitle{Blue Sky Ideas Track}
%\subtitle{Robotics Track}
%\subtitle{JAAMAS Track}
%\subtitle{Doctoral Mentoring Program}

%\subtitlenote{The full version of the author's guide is available as \texttt{acmart.pdf} document}

% AAMAS: submissions are anonymous for most tracks
%\author{Paper \#445}  % put your paper number here!

%% example of author block for camera ready version of accepted papers: don't use for anonymous submissions
%
\author{Hongyao Tang}
\affiliation{%
  \institution{Tianjin University}
}
\email{bluecontra@tju.edu.cn}

\author{Li Wang}
\affiliation{%
  \institution{Tianjin University}
}
\email{wangli@tju.edu.cn}

\author{Zan Wang}
\affiliation{%
  \institution{Tianjin University}
}
\email{wangzan@tju.edu.cn}

\author{Tim Baarslag}
\affiliation{%
  \institution{Centrum Wiskunde \& Informatica}
}
\email{tb1m13@ecs.soton.ac.uk}

\author{Jianye Hao}
\affiliation{%
  \institution{Tianjin University}
}
\email{jianye.hao@tju.edu.cn}
%\authornote{The secretary disavows any knowledge of this author's actions.}
%\affiliation{%
%  \institution{Institute for Clarity in Documentation}
%  \streetaddress{P.O. Box 1212}
%\email{webmaster@marysville-ohio.com}
%
%\author{Lars Th{\o}rv{\"a}ld}
%\authornote{This author is the
%  one who did all the really hard work.}
%\affiliation{%
%  \institution{The Th{\o}rv{\"a}ld Group}
%  \streetaddress{1 Th{\o}rv{\"a}ld Circle}
%\email{larst@affiliation.org}
%\affiliation{%
%  \institution{Inria Paris-Rocquencourt}
%\affiliation{%
% \institution{Rajiv Gandhi University}
%\affiliation{%
%  \institution{Tsinghua University}
%  \streetaddress{30 Shuangqing Rd}
%  \country{China}
%\affiliation{%
%  \institution{Palmer Research Laboratories}
%  \streetaddress{8600 Datapoint Drive}
%\affiliation{\institution{The Th{\o}rv{\"a}ld Group}}
%\email{jsmith@affiliation.org}
%
%\author{Julius P.~Kumquat}
%\affiliation{\institution{The Kumquat Consortium}}
%\email{jpkumquat@consortium.net}
%
%% The example's default list of authors is too long for headers

\begin{abstract}  % put your abstract here!
Multiagent coordination in cooperative multiagent systems (MASs) has been widely studied in both
fixed-agent repeated interaction setting and the static social learning framework.
However, two aspects of dynamics in real-world multiagent scenarios are currently missing in existing works.
First, the network topologies can be dynamic where agents may change their connections through rewiring during the course of interactions.
Second, the game matrix between each pair of agents may not be static and usually not known as a prior.
Both the network dynamic and game uncertainty increase the coordination difficulty among agents.
In this paper, we consider a multiagent dynamic social learning environment in which
each agent can choose to rewire potential partners and interact with randomly chosen neighbors in each round.
We propose an optimal rewiring strategy for agents to select most beneficial peers to interact with
for the purpose of maximizing the accumulated payoff in repeated interactions.
We empirically demonstrate
the effectiveness and robustness of our approach through comparing with benchmark strategies.
The performance of three representative learning strategies under our social learning framework with
our optimal rewiring is investigated as well.
\end{abstract}

% AAMAS: the ACM CCS are not needed within AAMAS papers
%%
%% The code below should be generated by the tool at
%% http://dl.acm.org/ccs.cfm
%\begin{CCSXML}
%<ccs2012>
%  <concept_id>10010520.10010553.10010562</concept_id>
%  <concept_desc>Computer systems organization~Embedded systems</concept_desc>
%  <concept_significance>500</concept_significance>
%  <concept_id>10010520.10010575.10010755</concept_id>
%  <concept_desc>Computer systems organization~Redundancy</concept_desc>
%  <concept_significance>300</concept_significance>
%  <concept_id>10010520.10010553.10010554</concept_id>
%  <concept_desc>Computer systems organization~Robotics</concept_desc>
%  <concept_significance>100</concept_significance>
%  <concept_id>10003033.10003083.10003095</concept_id>
%  <concept_desc>Networks~Network reliability</concept_desc>
%  <concept_significance>100</concept_significance>
%</ccs2012>
%\end{CCSXML}
%
%\ccsdesc[500]{Computer systems organization~Embedded systems}

  % put your semicolon-separated keywords here!

\maketitle

%%%%%%%%%%%%%%%%%%%%%%%%%%%%%%%%%%%%%%%%%%%%%%%%%%%%%%%%%%%%%%%%%%%%%%%%%%%%%%%%%%%%%%%%%%%%%%%%%%%%%%%%%
%% start of main body of paper

%%%%%%%%%%%%%%%%%%%%%%%%%%%%%%%%%%%%%%%%%%%%%%%%%%%%%%%%%%%%%%%%%%%%%%%%%%%%%%%%%%%%%%%%%%%%%%%%%%%%%%%%%
%% Introduction

\section{Introduction}

Multiagent coordination in cooperative multiagent systems (MASs) is a significant and widely studied problem
in the literature.
In cooperative MASs, the agents share common interests defined by the same reward functions \cite{sen2007emergence}.
This requires the agents to have the capability of coordinating with each other effectively towards desirable outcomes.

Until now, a lot of research efforts have been devoted to addressing multiagent coordination problems in
cooperative MASs \cite{claus1998dynamics,panait2005cooperative,matignon2012independent,lauer2000algorithm,
kapetanakis2002reinforcement,matignon2008study}.
One class of research focuses on coordination issues in fixed-agent repeated interaction settings.
Claus and Boutilier \shortcite{claus1998dynamics} firstly proposed two representative learning strategies - Independent Learner (IL) and Joint-Action Learner (JAL) to investigate their coordination performance in repeated two-agent cooperative games.
Later a number of improved strategies
\cite{lauer2000algorithm,kapetanakis2002reinforcement,panait2005cooperative,matignon2008study}
have been proposed to achieve coordination more efficiently and overcome the noise
introduced by the high mis-coordination cost and stochasticity of games.
Matignon et al. \shortcite{matignon2012independent} extensively investigate the comparative performance
of existing independent strategies in terms of their coordination efficiency in two-agent repeated cooperative games.
Their results show that perfect coordination can hardly be achieved for fully stochastic games.
However, in practical complex environments, the interactions between agents can be sparse,
i.e., it is highly likely that each agent may not have global access and only have the opportunity to
interact with local neighbors.
Moreover, agent's interacting partners are not fixed and may change frequently and randomly.
For example, in social media \cite{ellison2007social},
it is commonly accepted that the user attention span is limited,
hence it is of great importance to identify interactions that have optimal rewards.
Thereby, another line of research focuses on investigating the question of how a population of cooperative agents can effectively coordinate among each other under the social learning framework 
\cite{sen2007emergence,villatoro2011social,yu2013emergence,hao2013dynamics,hao2014reinforcement,hao2017dynamics,airiau2014emergence,mihaylov2014decentralized}.
For example,
Hao and Leung \shortcite{hao2013dynamics} extend IL and JAL into the social learning framework and find that better coordination performance can be achieved by leveraging the observation mechanism.

Most of existing works under the social learning framework assume that agents are located in a static network.
During each round of interaction, each agent plays the same cooperative game
with a randomly selected partners from its neighborhood.
However, two important aspects of dynamics in real-world multiagent scenarios are currently missing.
First, the cooperative games might be different for different pairs of agents due to a variety of reasons
(e.g., the difference of agents' capabilities and preferences, and the difference of interaction timing and locations)
in practical scenarios.
For example, in sensor networks \cite{zhang2013coordinating},
the overlapping scanning areas of each pair of sensors are of various importance
and different rewards are obtained when specific scanning areas are coordinated between two senor agents.
Another example is that in negotiation domain \cite{baarslag2015optimal},
an agent may have different preferences on each offer provided by different opponents.
Therefore, in this paper,
we relax this assumption by assuming that the payoff matrix between each pair of agents is different
which is generated from certain probability distribution,
and the payoff information is unknown before their interaction.

Second, the network topologies can be dynamic.
For example in social networks \cite{ellison2007benefits,kwak2010twitter},
it is common that users follow and unfollow other users
on their own initiative due to individual preference and interest.
Therefore, in this paper,
we consider a dynamic environment where each agent
is given the opportunities to change its connections through rewiring to interact with agents
which may bring higher payoffs during the course of interactions.
The rewiring mechanism has been previously used to explore indirect reciprocity \cite{peleteiro2014exploring} or cope with cheaters \cite{griffiths2010changing} to promote cooperations among agents in non-cooperative environments such as $Donation$ games.
In contrast, in this paper, we focus on cooperative environments
where agents can utilize the rewiring mechanism to increase their benefits in long-run interactions
by breaking connections to bad-performance neighbors and replacing them with new connections.

In this work, during each round, each agent goes through two main phases: the rewiring phase and the interaction phase.
In the rewiring phase, each agent is likely to be given the opportunity to alter their connections
through rewiring for higher payoffs in future interactions.
The key problem here is how to compute the optimal rewiring strategy
to maximize the long-run interaction payoff against the network dynamics.
To make efficient rewiring decisions, each agent has to take two aspects of uncertainties into consideration,
i.e., the uncertainties of opponents' behaviors and the payoff matrix with unknown peers.
To this end, we model an agent's rewiring problem as a sequential decision-making problem
and propose an optimal rewiring approach for agents to select most beneficial peers among all reachable agents.
In the interaction phase, agents learn their policies from pair-wise interactions
through playing certain cooperative game
against randomly selected opponent from its neighborhood.
Empirical results show that our optimal rewiring strategy outperforms other existing benchmark strategies in terms of agents' average accumulated payoff and robustness against different environments.
Besides, the relative performance of three representative learning strategies (i.e., Fictitious Play, Joint-Action Learner and Joint-Action WoLF-PHC) is analyzed as well.

The remainder of the paper is organized as follows.
In Section 2,
we give an overview of related works.
In Section 3,
we give a formal description of our coordination problem in dynamic cooperative MASs.
In Section 3,
the social learning framework and both rewiring and learning strategies are described.
In Section 4,
we empirically demonstrate the efficiency of our rewiring approach and analyze the influence of three learning strategies
under the social learning framework with rewiring.
Lastly conclusion and future work are given in Section 5.

%%%%%%%%%%%%%%%%%%%%%%%%%%%%%%%%%%%%%%%%%%%%%%%%%%%%%%%%%%%%%%%%%%%%%%%%%%%%%%%%%%%%%%%%%%%%%%%%%%%%%%%%%
%% Related work

\section{Related Work}
There has been a large amount of research in the multiagent reinforcement learning literature
on solving coordination problem in cooperative MASs
%\cite{claus1998dynamics,panait2006lenient,matignon2012independent,hao2013dynamics,hao2014reinforcement,hao2015multiagent,
%lauer2000algorithm,kapetanakis2002reinforcement,panait2005cooperative,matignon2008study}.
One line of research focuses on solving coordination problem in fixed-agent repeated interaction settings
\cite{claus1998dynamics,lauer2000algorithm,kapetanakis2002reinforcement,panait2005cooperative,matignon2008study,panait2006lenient,matignon2012independent}.
Claus and Boutilier \shortcite{claus1998dynamics} investigate the coordination performance of
the Independent Learner and the Joint-Action learner,
in the context of repeated two-agent cooperative games.
It shows that both types of learners achieve success
in simple cooperative games.
Lauer and Riedmiller \cite{lauer2000algorithm} propose the distributed Q-learning algorithm base on the optimistic assumption.
It is proved that optimal joint actions can be guaranteed to
achieve if the cooperative game is deterministic.
FMQ heuristic \cite{kapetanakis2002reinforcement} and recursive FMQ \cite{matignon2008study}
are proposed to alter the Q-value estimation function to handle the stochasticity of the games.
FMQ obtains success in partially stochastic climbing games but still can not
achieve satisfactory performance in fully stochastic cooperative games.
Panait et al. \cite{panait2006lenient} propose the lenient multiagent learning
algorithm to overcome the noise introduced by the stochasticity of the games.
The results show that coordination on the optimal joint action
can be achieved in more than 93.5\% of the runs in the fully stochastic climbing game.
Matignon et al. \shortcite{matignon2012independent} review all existing independent
multiagent reinforcement learning algorithms and evaluate and discuss their strength and weakness.
It shows that all of them fail to achieve perfect coordination for fully stochastic games
and only recursive FMQ can achieve coordination for 58\% of the runs.

The other line of research considers the multiagent coordination problem under the social learning framework
\cite{sen2007emergence,villatoro2011social,yu2013emergence,hao2013dynamics,hao2014reinforcement,
hao2017dynamics,airiau2014emergence,mihaylov2014decentralized},
where agents learn through pair-wise interactions by playing the same game.
Sen and Airiau \cite{sen2007emergence} propose the social learning model and investigate
the emergence of consistent coordination policies (norms) in MASs (e.g., conflicting-interest games) under
this social learning framework.
Hao and Leung \shortcite{hao2013dynamics} investigate the multiagent coordination problems
in cooperative environments under the social learning framework.
They demonstrate individual action learners (IALs) and joint action learners (JALs)
and achieve better coordination performance by leveraging the observation mechanism.
Most of previous works assume that agents learn through playing the same cooperative game
with randomly selected partners from their neighborhood within a static network.
In contrast, in this paper, we consider a dynamic environment in which
agents learn through pair-wise interactions
through playing uniquely and randomly generated cooperative games.

There are some existing works on using rewiring mechanism to support cooperation in social networks
\cite{peleteiro2014exploring,griffiths2010changing,hales2005applying,dekker2007realistic}.
Peleteiro et al. \cite{peleteiro2014exploring} propose a new mechanism based on three main
pillars: indirect reciprocity, coalitions and rewiring,
to improve cooperation among self-interested agents placed in a complex network.
Rewiring, as a part of the proposed mechanism, alters the worst social links
with the best coalition members  the neighbors' reputation coalitions information.
They confirm that the use of rewiring indeed improves cooperation
when they play the donation game in their social scenario.
Griffiths and Luck \shortcite{griffiths2010changing} present and demonstrate a decentralised tag-based mechanism
to support cooperation in the presence of cheaters without requiring reciprocity.
The simple rewiring enables agents to change their neighbour connections, in particular by removing connections to
the worst neighbours and replacing them with connections to neighbours with whom others have had positive experiences.
Their results show that cooperation can be improved by enabling agents to change their neighbour connections.
In contrast, in this paper, we consider the multiagent coordination problem in cooperative MASs
and how rewiring can increase the benefits of individual agents during long-run interactions under the social learning framework.

%%%%%%%%%%%%%%%%%%%%%%%%%%%%%%%%%%%%%%%%%%%%%%%%%%%%%%%%%%%%%%%%%%%%%%%%%%%%%%%%%%%%%%%%%%%%%%%%%%%%%%%%%
%% Problem description

\section{Problem Description}
We consider a population $N$ of agents,
each of which has (bidirectional) connections to its neighbors.
We use $O_i$ to denote agent $i$'s neighborhood,
such that each agent $i$ is only able to interact with its neighbors in $O_i$.
We model the strategic interaction among each pair of agents as a cooperative game,
where each agent always receives the same payoff under the same outcome.
One example of deterministic two-action cooperative games is shown in Fig.1(a),
in which there exist two optimal Nash equilibria.
In this game, agents are desired to coordinate their actions towards a consistent Nash equilibrium to maximize their benefits.

Previous works
\cite{panait2006lenient,matignon2012independent,hao2013dynamics,hao2014reinforcement,hao2015multiagent}
have investigated the problem of how a population of agents could achieve coordination
under the same cooperative environments (games) to maximize the system's overall benefits.
However, in general, the payoff matrix between any pair of agents may be different depending on
the agents' interacting environments.
Following the setting in \cite{damer2008achieving},
we assume that each game is drawn independently from certain probability distribution
to model agents' interaction environments in a generalized way.
Without loss of generality, a general form of 2-action cooperative game $G_i^j$
between agent $i$ and $j$ is shown in Fig.1(b),
where the value of $u_a$ (or $u_b$) is sampled from a stochastic variable $x_a$ (or $x_b$)
following certain cumulative probability distribution $F_a(x)$ (or $F_b(x)$),
and $\alpha < x_a (x_b)$ represents the payoff under mis-coordination.
The payoff matrix of any pair of agents is not known as a prior, which can be learned through repeated interactions.
In our settings, we assume that each agent in the environment can observe
the actions of its interaction neighbor at the end of each interaction.

Inspired from human society \cite{ellison2007benefits,kwak2010twitter}, agents should have the freedom of choosing whom they want to interact with.
Thus, in this paper, we consider that the underlying interaction topology is dynamic:
agents are allowed to break existing connections which they do not benefit from
and establish new connections with non-neighbor agents through rewiring.
Each rewiring is associate with certain cost and we use $c_i^j$
to denote the rewiring cost of agent $i$ when it establishes new connection with another agent $j$.
As in many scenarios, agents usually do not have access to all agents in the environment but only local access
due to physical limitations, e.g. the distance of signal reception and delivery,
the radius of human relation circle \cite{crow1997ieee,kwak2010twitter}.
Hence it is not feasible to allow each agent to be able to rewire any other agents in the environments,
and here we assume that each agent is only allowed to establish connections with the set of reachable peers,
which can be defined as the set of agents within certain distance of agent $i$.
We use $\{O_i \cup \bar O_i\}$ to denote agent $i$'s reachable peers,
consisting of $O_i$ for agent $i$'s neighborhood
and $\bar O_i$ for all non-neighbor agents that agent $i$ can potentially interact with through rewiring.

Moreover, unlimited increase of communication in real-world networks is usually impractical
and the number of connections should have a upper bound due to finite connection resources \cite{zhang2013coordinating}.
Each agent should break an old connection each time it decides to establish a new connection through rewiring.\footnote{This can be easily
extended to the case of allowing multiple rewiring by ranking the expected payoff of reachable peers in the descending order.}
The interest of any rational agent is to maximize its accumulative payoff during the course of interactions.
Therefore, a rational agent needs to balance the tradeoff between rewiring to explore more beneficial partners to interact with
and exploiting the current connected neighbors to avoid rewiring cost.

\section{Social Learning Framework}
Under the social learning framework, there is a population $N$ of agents and each agent learns its policy through
repeated pair-wise interactions with its neighbors.
The initial neighborhood of each agent is determined by the underlying topology and three representative topologies are considered here: Regular Network, Small-world Network, Scale-Free Network \cite{watts1998collective,albert2002statistical}.
The overall interaction flow under the social learning framework is shown in Algorithm 1.

\begin{figure}
\centering

\subfigure[]{
\label{sub1}
\includegraphics[width=0.15\textwidth]{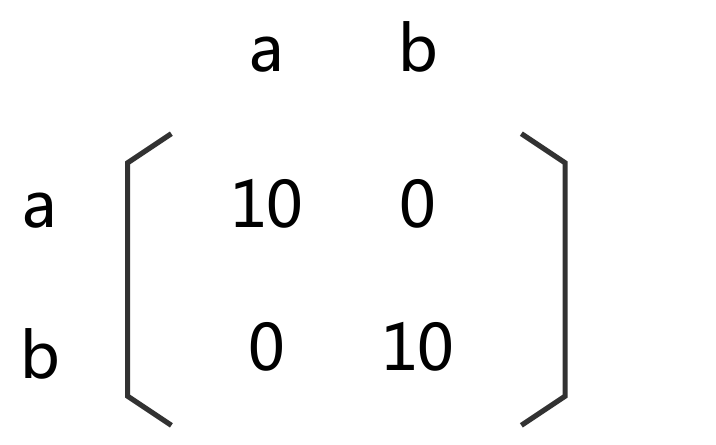}}
\hspace{1cm}
\subfigure[]{
\label{sub2}
\includegraphics[width=0.15\textwidth]{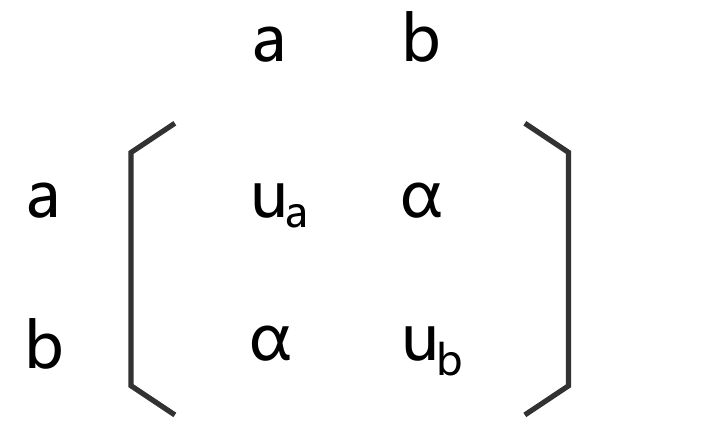}}

\caption{$(a)$ An instance of cooperative games. $(b)$ A general form of a cooperative game.}
\label{fig:1}
\end{figure}

\begin{algorithm}
  \caption{Overall interaction protocol under the social learning framework for each agent $i \in N$}
  \begin{algorithmic}[1]
	\For{a number of interaction rounds}
	
		\If { random variable $p \le \varphi$}
			\State Perform rewiring action.
		\EndIf
		\State Play game $G_i^j$ with randomly chosen player $j \in O_i$.
		\State Obtain payoff and update its policy.
		\State Update neighbor $j$'s action model.
	\EndFor
    \label{code:Algorithm 1}
  \end{algorithmic}
\end{algorithm}

During each round, each agent $i$ first is given the opportunity of rewiring with probability $\varphi$.
During the rewiring phase, agent $i$ breaks its connection with a neighbor with poor-performance and establishes a new connection with another agent from its potential partners $\bar O_i$ (Line 2-4).
Next, in interaction phase, agent $i$ interacts with a randomly selected neighbor agent $j$ from its neighborhood $O_i$
by playing their corresponding cooperative game $G_i^j$ (Line 5).
After the interaction, agent $i$ and $j$ receive the corresponding payoff and update their policies and opponents' models respectively (Line 6-7).
The details of the rewiring and learning strategies will be introduced in following subsection.

\subsection{An Optimal Rewiring Strategy}
The goal of rewiring is to explore the set of unconnected peers (potential partners $\bar O$)
in the population and seek more beneficial partners to interact with.
Each agent is faced with two aspects of uncertainties:
the uncertainty of the behaviors of its neighbors and unknown peers;
the uncertainty of the payoff matrix during interaction with unknown peers.
To select the optimal agent to rewire, we need to evaluate the potential benefits of
interacting with agent $j$
by taking the aforementioned uncertainties into consideration.

Before making a rewiring decision, each agent first needs to evaluate the benefits of interacting with each neighbor.
One natural way is to use the optimistic assumption, i.e., computing the expected payoff of selecting different actions
based on the estimated policy of a neighbor and picking the action with the highest expected value.
In other words,
agent $i$ evaluates the expected payoff $v_i^j$ of interacting with agent $j \in O_i$
by using the highest expected payoff that can be received among all possible actions $A_i$.
Formally we have:
\begin{equation}
  v_i^j = \max_{m \in A_j}p_i^j(m)u_m + \left(1 - p_i^j(m)\right)\alpha,
\end{equation}
where $u_m$ represents the explicit payoff when both players choose action $m$ by sampling $x_m$ from $F_m(x)$ during historic interactions,
and $\alpha$ is for the mis-coordination payoff.
$p_i^j(m)$ denotes agent $i$'s estimated probability of agent $j$ choosing action $m$. The value of $p_i^j(m)$ can be easily obtained as the empirical frequency distribution of agent $j$'s actions based on historical interactions
and other advanced techniques such as weighting more on recent experiences may also be considered.
Provided there is no need to rewire,
we use agent $i$'s worst-case expected payoff among all the neighbors as its baseline utility. Formally we have:
\begin{equation}
  v_i^*(O_i) = \min_{j \in O_i}v_i^j.
\end{equation}

For each unknown potential partner $j \in \bar O_i$ to interact with, agent $i$ is faced with two aspects of uncertainties: agent $j$'s behavior, and the payoff matrix $G_i^j$.
For a previously unseen partner $j$'s behavior, it can be estimated
using the observed expected behaviors from agent $i$'s existing neighborhood, while the payoff matrix $G_i^j$ has to be learned through repeated interactions after agent $i$ and $j$ are connected through rewiring.
Therefore, agent $i$'s estimation of expected benefit $x_i^j$ from interacting with an unknown agent $j$ can be formalized as follows:
\begin{equation}
  x_i^j = \max_{m \in A_j}\bar p_i^j(m)x_m + \left(1 - \bar p_i^j(m)\right)\alpha,
\end{equation}
where $x_i^j$ is a stochastic variable and $F_j^i(x)$ denotes the corresponding cumulative distribution function of $x_i^j$,
and $\alpha$ is the payoff under mis-coordination.
$\bar{p}_i^j(m)$ is agent $i$'s estimated probability of agent $j$ choosing action $m$ within its neighborhood.
As no interaction has been made between agent $i$ and the unknown peer $j \in \bar O_i$ before,
the distribution over agent $j$'s actions can not be observed a prior
but estimated from the neighbors which once interact with agent $j$.

Intuitively, it is reasonable for agent $i$ to establish new connection with any unknown potential partner $j$
if rewiring leads to higher expected payoff.
The difference between expected payoff $x_i^j$ and baseline value causes
a incoming benefit which is promising to pay back the rewiring cost $c_i^j$ and to generate better payoffs in further interactions.
Otherwise, agent $i$ should keep its current neighborhood unchanged.
During each rewiring phase, each agent $i$ has to unlink a bad-performance neighbor first before rewiring.
Note that there is no need for agent $i$ to consider those agents already disconnected during previous rewiring,
since the expected payoff of interacting with discarded agent $j$ is obviously below the current interaction baseline,
i.e. $v_i^j < v_i^*(O_i)$.
%agent $i$ is pretty sure that the expected payoff obtained through interacting with those agents are worse than interacting with current neighbors.

Each agent's situated environments are continuously changing as it breaks old connections and establishes new ones, which also influences the following interactions and rewiring decisions thereafter. Therefore, each agent is actually faced with a sequential rewiring decision-making problem.
Formally, we propose modeling the above sequential rewiring problem for each agent $i$ as an Markov Decision Process (MDP), $M_i=<S_i,A_i,P_i,R_i>$ as follows:
\begin{itemize}
\item A finite set of states $S_i$: each state $\epsilon$ of agent $i$
can be represented as a tuple $<\bar O_i, y_i>$,
in which $\bar O_i$ describes the set of potential partners of agent $i$ and
$y_i = v_i^*(O_i)$ is agent $i$'s current baseline value.
\item A finite set of actions $A_i$:
agent $i$'s action set under $\epsilon$ is $A_i(\epsilon) = \{\bar O_i, \textbf{Null}\}$ which
consists of actions of rewiring with an agent in $\bar O_i$ and the \textbf{Null} action for not rewiring.
\item A transition function $P_i$:
$P_i(\epsilon,a,\epsilon')$ represents the probability of reaching state $\epsilon'$
after taking action $a$ under state $\epsilon$.
The transition is probabilistic because the value of baseline $y_i$ may change stochastically.
If agent $i$ rewires agent $j \in \bar O_i$,
the new baseline $y_i' = v_i^*({O_i}')$ with the updated ${O_i}'$ is determined
by new included neighbor $j$ and left neighbors after breaking the worst-performing connection.
%The expected payoff $x_i^j$ is unknown as a prior and can be estimated by its distribution $F_j^i(x)$.
\item A reward function $R_i$: $R_i(\epsilon,a)$ denotes the reward of executing action $a$ under state $\epsilon$, e.g., $R_i(\epsilon) = -c_i^j$ if agent $i$ chooses to rewire agent $j$ and $R_i(\epsilon) = 0$ if stops rewiring.
\end{itemize}

Under the above MDP formulation, our goal is to construct a rewiring strategy $\pi_i(\epsilon)$ specifying which new connection to establish,
or to not rewire under each state $\epsilon$.
We start with a short-sight version where each agent is only interested in maximizing its one-shot payoff after rewiring.
The utility $U(\pi_i,\epsilon)$ of a policy $\pi_i$ can be defined recursively as follows,

\begin{equation}
    U(\pi_i,\epsilon) =
    \left\{
                 \begin{array}{lr}
                 \int_{-\infty}^\infty U(\pi_i, \epsilon') \;\text{d}F_{x_{\pi_i(\epsilon)}}^i(x)
			 & \text{if} \ \pi_i(\epsilon) \in \bar O_i, \\
                 -c_i^{\pi_i(\epsilon)} \\
			 y_i & \text{otherwise stop.}
                 \end{array}
    \right.
\end{equation}
where $\epsilon' = <\bar O_i \backslash \{\pi_i(\epsilon)\},y_i'\}>$ denotes the new rewiring state after
agent $\pi_i(\epsilon)$ is rewired,
%payoff matrix $G_i^j$ has been observed,
where $y_i'$ is the new expected baseline payoff as the neighborhood $O_i$ is altered.

Our final goal is to find an optimal policy $\pi_i^* = arg\max_{\pi_i} U(\pi_i,\epsilon)$ among all feasible policies,
which considers the following: either not rewire and obtain baseline $y_i$, or rewire an unknown partner $j \in \bar O_i$ by sampling $x$ from $x_i^j$ at cost $c_i^j$, while taking the following into account:

\begin{itemize}
\item $x \le y_i$: this indicates the expected payoff interacting with agent $j$ is not better than its current baseline and $x$ should be the new baseline value after rewiring. Thus the expected utility under state $<\bar O_i, y_i>$ becomes $-c_i^j + U(\pi_i^*,<\bar O_i \backslash \{j\},x>)$;
\item $x >y_i$: the expected payoff interacting with agent $j$ is higher than our baseline value. In this case, the expected utility under state $<\bar O_i, y_i>$ becomes
$-c_i^j + U(\pi_i^*,<\bar O_i \backslash \{j\},y_i'>)$, where $y_i^\prime$ is the updated baseline value since agent $i$ breaks the connection with the worst-performance neighbor during rewiring.
\end{itemize}
%\begin{quote}
%1) $x \le y_i$: this indicates the expected payoff interacting with agent $j$ is not better than its current baseline and $x$ should be the new baseline value after rewiring. Thus the expected utility under state $<\bar O_i, y_i>$ becomes $-c_i^j + U(\pi_i^*,<\bar O_i \backslash \{j\},x>)$;
%\end{quote}
%\begin{quote}
%2) $x >y_i$: the expected payoff interacting with agent $j$ is higher than our baseline value. In this case, the expected utility under state $<\bar O_i, y_i>$ becomes
%$-c_i^j + U(\pi_i^*,<\bar O_i \backslash \{j\},y_i'>)$, where $y_i^\prime$ is the updated baseline value since agent $i$ breaks the connection with the worst-performance neighbor during rewiring.
%\end{quote}

The above way of formalizing the dynamics of an optimal strategy only considers the one-shot interaction benefits.
However, in our social learning framework, each agent tries
to maximize accumulated payoffs during multi-rounds interactions.
To this end, we propose a new far-sight way of modeling our rewiring problem such that each agent
considers the accumulated expected payoff through multiple interactions with any neighbor after rewiring.
Formally, the $K$-step utility function $U_K(\pi_i^*,\epsilon)$ of an optimal strategy $\pi_i^*$ must satisfy the following recursive relation:

\begin{equation}
\begin{aligned}
\ U_K(\pi_i^*,\epsilon) = & \min \{
	K y_i, \max_{j \in \bar O_i} \{
  		-c_i^j  +  \ U_K(\pi_i^*,<\bar O_i \backslash \{j\},x>) \cdot F_j^i(y_i) \\
		& +  \ \int_{x=y_i}^\infty U_K(\pi_i^*,<\bar O_i \backslash \{j\},y_i'>) \cdot \;\text{d}F_j^i(x)
  \}
\},
\end{aligned}
\end{equation}
where the rewiring sight value of $K$ models an agent's far-sight degree
by taking the accumulated payoff from $K$ rounds of interactions into account during each rewiring.
We can see that Eq.5 is essentially a Bellman equation, which could be solved by backward induction in principle.
However, even for a moderate-size MAS, this approach quickly becomes intractable.
To this end, we provide a simple but optimal method to compute which agents to rewire or not in $O(n^2)$ time (further explained later), which is inspired from Pandora's Rule\footnote{Pandora's Rule \cite{weitzman1979optimal}
 is a solution concept for interaction under uncertainty; this framework has a wide range of applications to
 dynamic alternative selection problems, such as optimal service provider selection in Task-Procurement Problem \cite{gerding2009mechanism},
 data management \cite{baarslag2017permission}  and optimal preference elicitation in negotiation  \cite{baarslag2015optimal}.}
\cite{weitzman1979optimal}.

Supposing that
agent $i$ has current baseline $y_i$ and a potential partner $j$ in $\bar O_i$,
to determine whether to rewire partner $j$ through our $K$-sight optimal rewiring approach,
we have:

\begin{itemize}
\item In the following $K$ rounds, the expected interaction value we could obtain from neighborhood $O_i$ is at least:
\begin{equation}
	K y_i.
\end{equation}

\item If choose to rewire agent $j$, we could get a net benefit of the following $K$-rounds interactions in the worst cases as follows:
\begin{equation}
	-c_i^j + K (\int_{-\infty}^{y_i}x \ \;\text{d}F_j^i(x) + {y_i}' \int_{y_i}^{\infty}\;\text{d}F_j^i(x) ).
\end{equation}
where ${y_i}'$ is the new baseline if $x \ge y_i$ after breaking the connection with old baseline one.
We define ${y_i}'$ as ${y_i}' = \min\{x, y_i^{sec}\}$ where $y_i^{sec}$ is the second minimum expected payoff in $O_i$
%or the updated baseline value excluding $x$,
given that agent $i$ has at least two neighbors before rewiring.\footnote{If agent $i$ has only one neighbor, the new baseline degenerates to be $y_i' = x$, which is similar and simpler.}
Formally we have:
\begin{equation}
	{y_i}' = \int_{-\infty}^{y_i^{sec}}x \ \;\text{d}F_j^i(x) + y_i^{sec} \int_{y_i^{sec}}^{\infty}\;\text{d}F_j^i(x).
\end{equation}

\item Further, we use $\Lambda$ for the single-round payoff difference between not rewiring (Eq.6) and rewiring agent $j$ (Eq.7) which can be formalized as follows:
%The benefit difference $\Lambda$ can be formalized as follows:
\begin{equation}
	\Lambda = \int_{-\infty}^{y_i}x \ \;\text{d}F_j^i(x) + {y_i}' \int_{y_i}^{\infty}\;\text{d}F_j^i(x) - y_i -c_i^j/K.
\end{equation}
If $\Lambda = 0$, agent $i$ is just indifferent between rewiring $j$ and not rewiring.
Otherwise, the larger $\Lambda$ value indicates that agent $j$ is more beneficial to rewire.
\end{itemize}

For each unknown partner $j$, the expected payoff satisfies the cumulative distribution function $F_j^i(x)$ and the rewiring cost is $c_i^j$.
We can calculate an index $\Lambda_i^j$ through Eq.9,
which fully captures the relevant information about agent $j$:
it should be rewired when it has the highest positive index
and exceeds the interaction baseline of current neighborhood. It is proven
in \cite{weitzman1979optimal} that this strategy is optimal in terms of
expected reward Eq.5.

The overall K-sight rewiring strategy is shown in Algorithm 2.
At each rewiring phase, each agent $i$
first calculates the baseline interaction value $y_i$ and second minimum expected value $y_i^{sec}$ in $O_i$ (Line 1-6).
Second, for each potential partner $w \in \bar O_i$,
its corresponding index $\Lambda_i^w$ is computed following Eq.9 (Line 7-9).
Finally, agent $i$ makes the rewiring decision accordingly ---
to rewire target agent $\tau$ with the maximum value of $\Lambda_{max}$ at cost $c_i^{\tau}$ if $\Lambda_{max} \ge 0$ or not to rewire otherwise (Line 10-15).
Agent $i$ unlinks the worst-performing neighbor $k$ before rewiring a new partner:
\begin{equation}
  k = arg \min_{j \in O_i}v_i^j.
\end{equation}
After certain rounds, each agent stops rewiring and converges to an optimal neighborhood.

\begin{algorithm}[H]
  \caption{K-Sight rewiring strategy for agent $i$ with sight $K$ in each rewiring phase}
  \begin{algorithmic}[1]
	\For{each $j \in O_i$}
		\State $v_i^j = calculateExpectedReward(j)$;
	\EndFor
	\State $y_i = \min_{j \in O_i} v_i^j$;
	\State $k = \arg \min_{j \in O_i} v_i^j$;
	\State $y_i^{sec} = \min_{j \in O_i\backslash \{k\}} v_i^j$;
	\For{each $w \in \bar O_i$}
		\State $\Lambda_i^w = calculateK\_SightIndex(w,K,y_i,y_i^{sec})$;
	\EndFor
	\State $\Lambda_{max} = \max_{w \in \bar O_i} \Lambda_i^w$;
	\If {$\Lambda_{max} \ge 0$}
		\State $\tau = \arg \max_{w \in \bar O_i} \Lambda_i^w$;
		\State $UnlinkNeighbour(k)$;
		\State $RewiringTarget(\tau)$;
	\EndIf

    \label{code:Algorithm 1}
  \end{algorithmic}
\end{algorithm}

Next we analyze the computational complexity of our proposed algorithm.
In Algorithm 2, for each rewiring phase,
agent $i$ computes the interaction baselines of current neighborhood $O_i$ (Line 1-6) and the $\Lambda$ value (Line 6-9)
for each potential peer in $\bar O_i$ to find the optimal action (Line 10-15).
This leads to a computational complexity of $O(n)$
and the value of $n$ is for the size of reachable peers $\{O_i \cup \bar O_i\}$.
Solving the Bellman equation (Eq.5) is equivalent with calculating the optimal rewiring action
for all feasible states which are proportional to the size of potential peers.
Thus the total computational complexity of solving our sequential decision-making problem is $O(n^2)$.
Note that the computational complexity $O(n^2)$ can be quite low even when the population size $N$ increases significantly
since the number of reachable peers for each agent
is usually much smaller than the number of agents in whole populations, i.e. $n = |\bar O_i| \ll |N|$.

After rewiring phase, each agent proceeds to interact with an agent randomly selected from its neighborhood. We consider the following three representative learning strategies in the literature as agents' learning strategies.

\begin{itemize}
\item $\emph{Fictitious play(FP)}$.
FP is a well-known learning approach in literature.
Agent $i$ keeps a frequency count of agent $j$'s decisions from a
history of past moves and assumes that the opponent is playing a mixed strategy represented by this frequency distribution.
Agent $i$ chooses the best-response action $m$ to that mixed strategy to maximize its expected payoff.
Formally we have:
\begin{equation}
	m = \arg\max_{m \in A_j}p_i^j(m)u_m + \left(1 - p_i^j(m)\right)\alpha,
\end{equation}
where $p_i^j(m)$ is agent $i$'s estimated distribution over agent $j$'s actions and $u_m$ is the payoff
when both agent $i$ and $j$ choose action $m$, exactly the same in Eq.1.

\item $\emph{Joint Action Learner (JAL)}$ \cite{claus1998dynamics}.
JAL is a classic Multiagent Reinforcement Learning algorithm under the assumptions that each agent can observe the
actions of other agents.
A JAL agent learns its Q-values for joint actions.
Formally, $Q_i^j(a,b)$ represents agent $i$'s Q-value when agent $i$ and $j$ select action $a$ and $b$ respectively.
To determine the relative values of their individual actions,
agent $i$ assumes that each other agent $j$ will choose actions in accordance with agent $i$'s empirical
distribution over agent $j$'s action.
In general, agent $i$ assesses the expected value of each action $a$ as follows:
\begin{equation}
	EV(a) = \sum_{m \in A_j} Q_i^j(a,m) \ p_i^j(m),
\end{equation}
where $p_i^j(m)$ is agent $i$'s estimation for the probability of agent $j$ choosing action $m$.
%In JAL, we can use Q-values for approximation of explicit payoff $u$ in Eq.1.

\item $\emph{Joint-Action WoLF-PHC (JA-WoLF)}$.
WoLF \cite{bowling2001rational} extends Q-learning and learns mixed strategies with the idea of
quickly adapting when losing but being cautious when winning.
Specifically we modify WoLF to a Joint-Action version as agents can observe others' actions in our environments.
We replace $Q(s,a)$ with $Q(s,\vec{a})$ and each agent updates its Q-values for each joint-action.
In addition, to determine 'win' or 'lose', we need to keep the frequency distribution over opponents' actions.,
and agent $i$ adjust its learning rate $\delta$ against agent $j$ as follows:
\begin{equation}
	\delta =
    \left\{
                 \begin{array}{lr}
                 \delta_w & \text{if} \ \sum_{a \in A_i}\pi_i^j(a) \sum_{m \in A_j}Q_i^j(a,m)p_i^j(m) \\
			 & \ge \sum_{a \in A_i}\bar \pi_i^j(a) \sum_{m \in A_j}Q_i^j(a,m)p_i^j(m), \\
                 \delta_l
			 & \text{otherwise.}
                 \end{array}
    \right.
\end{equation}
where $\delta_w$ and $\delta_l$ denotes the learning rate when win or lose separately, $\pi_i^j$
is agent $i$'s strategy and $\bar \pi_i^j$ is the average strategy,
and $p_i^j(m)$ is the probability agent $i$ maintains for interacting with agent $j$.
\end{itemize}

%%%%%%%%%%%%%%%%%%%%%%%%%%%%%%%%%%%%%%%%%%%%%%%%%%%%%%%%%%%%%%%%%%%%%%%%%%%%%%%%%%%%%%%%%%%%%%%%%%%%%%%%%
% Experiments

\section{Experimental Evaluations}
In this section, we evaluate the performance of our optimal rewiring strategy with two benchmark
strategies against various interaction environments.
Following that, we analyze the performance of different learning strategies under our social learning framework.

\subsection{Parameter Settings}
We consider a population of hundreds of agents
and three representative network topologies: regular network, small-world network and scale-free network.
In our experiments, there is no apparent discrimination in results of these three initial topologies,\footnote{The network topologies of
interaction environments are changed along with the occurrence of dynamic rewirings. Thus,
different initial topologies show similar results in long-term interactions.}
and we use regular networks as default choice for following illustration.
Besides, we consider a wide range of interaction environments $(x,y,z)$
from three major aspects: the number of agents $x$, the size of each agent's initialized neighborhood $y$
and the size of each agent's reachable agents $z$.
For example, interaction environment $(100,4,12)$ denotes a scenario consisting of 100 agents which each agent
has 4 neighbors ($O$) and 12 reachable agents initially($O\cup \bar O$).
In following experiments the value of neighborhood $y$ are initialized
to be different constances equally for each agent.
Note that we put no limitation on the size of neighborhood and it can be constance or other function forms
to model the variance of agents' connection capabilities.
The set of reachable agents can also be defined in different ways.
For example, in \cite{peleteiro2014exploring} 
agents are allowed to do the rewiring through leaving its worst neighbor
and joins the one with the highest score in its coalition.
In contrast, in \cite{griffiths2010changing} an agent replaces its bad-performance neighbors and replaces them with
specific agents from the neighborhood of its (best) neighbors.
We set the value of reachable agents size $z$ to constances for the purpose of generalization.

For each pair of agents $i$ and $j$, the cooperative game $G_i^j$ is uniquely generated.
The payoffs $u_a$, $u_b$ on the diagonal of $G_i^j$
are sampled independently from $x_a$, $x_b$ according to probability distribution $F_a(x)$, $F_b(x)$
which are separately set to either a uniform distribution $U(n, m)$ or a beta distribution $Beta(\alpha,\beta)$,
with $n \ \textless \ m$ both uniformly sampled from $U(0, 1)$
and $\alpha$, $\beta \in \{1, . . . , 10\}$.
The mis-coordination payoff $\alpha$ is set to a constant value randomly sampled from the range of $[-0.2,0]$ for each payoff matrix.

%\begin{figure}
%\includegraphics[width=0.3\textwidth]{(100,4,12)-Ksight-40-250.pdf}
%\caption{Average payoffs of optimal rewiring strategy against varying rewiring sight and rewiring costs over 100 randomly initialized (100,4,12) networks with $\varphi = 0.01$.}
%\label{fig:2}
%\end{figure}

\subsection{Influence of Rewiring Sight}
In our optimal rewiring approach, the value of rewiring sight is the key parameter which may directly influence the performance.
We investigate the performance of varying sight values of $K$ with several rewiring costs.\footnote{The rewiring costs are usually set to be larger than single-round payoff because changing topologies is a non-trivial task.}
For learning strategy, we use FP for illustration and similar results can be observed for other learning strategies (i.e., JAL or JA-WoLF), which are omitted due to space limitation.

\begin{figure*}[h]
\centering

\subfigure[varying rewiring sight]{
\label{sub1}
\includegraphics[width=0.26\textwidth]{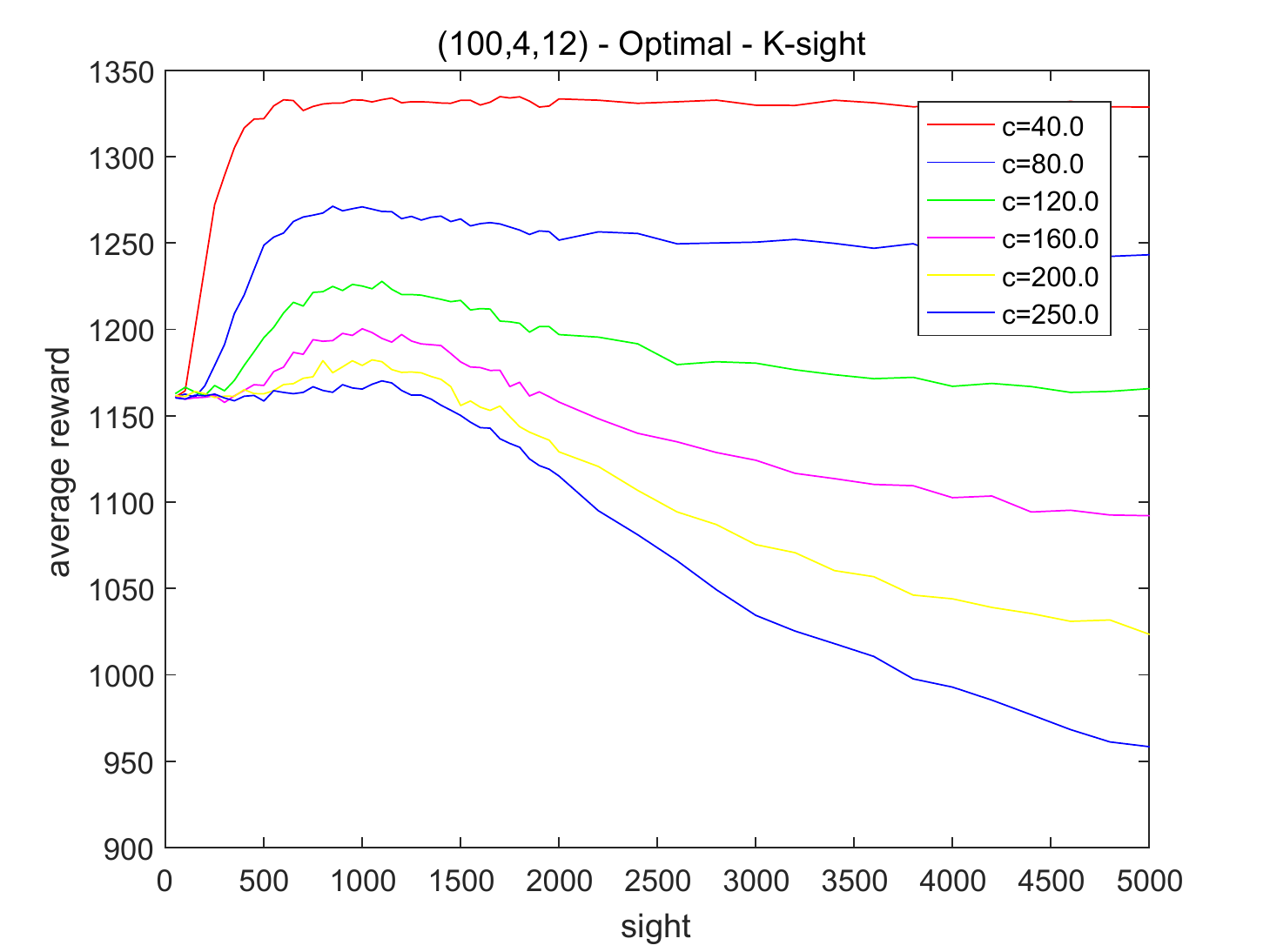}}
\hspace{-0.58cm}
\subfigure[in pure environments]{
\label{sub2}
\includegraphics[width=0.26\textwidth]{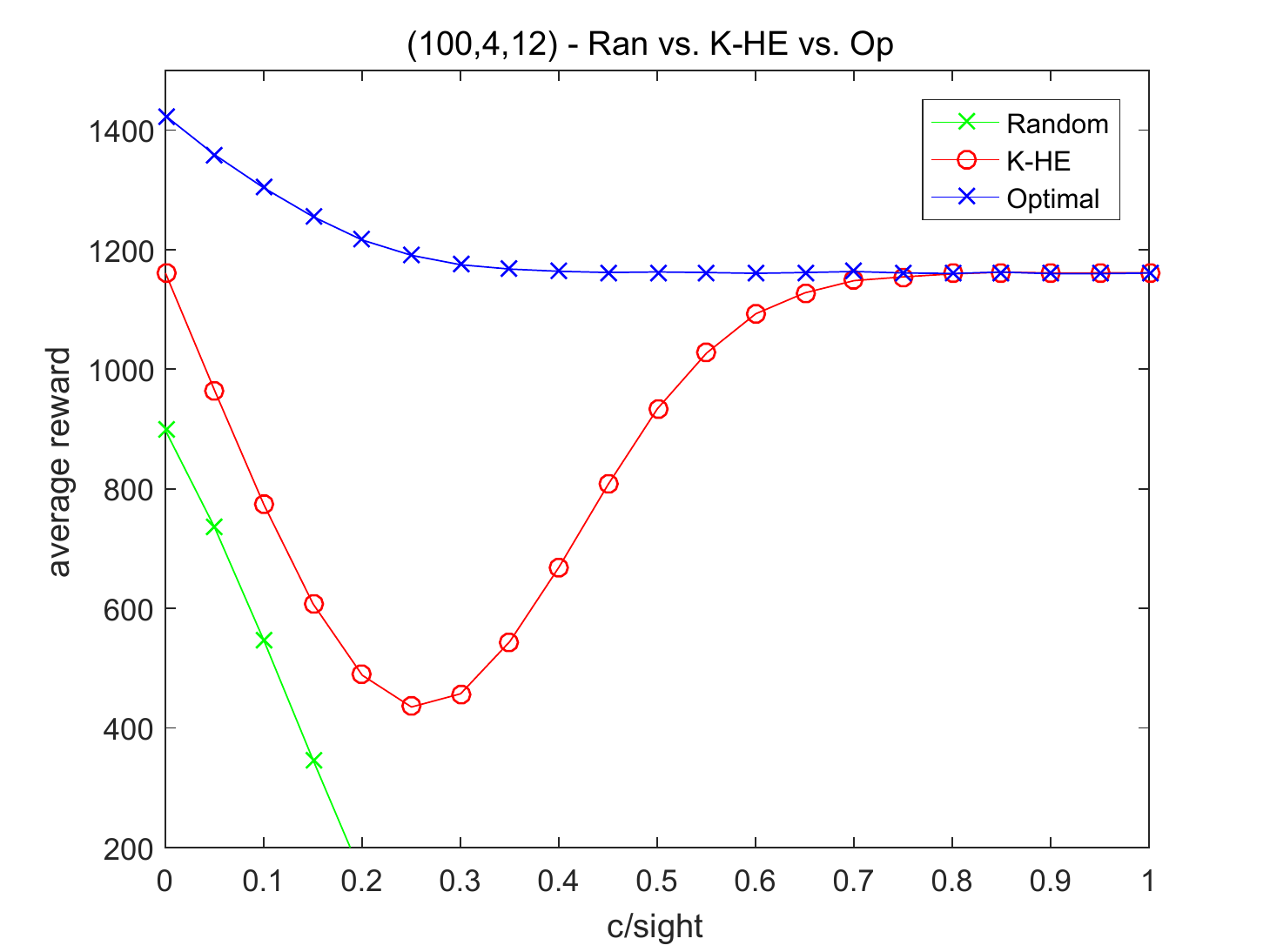}}
\hspace{-0.60cm}
\subfigure[in competitive environments]{
\label{sub3}
\includegraphics[width=0.26\textwidth]{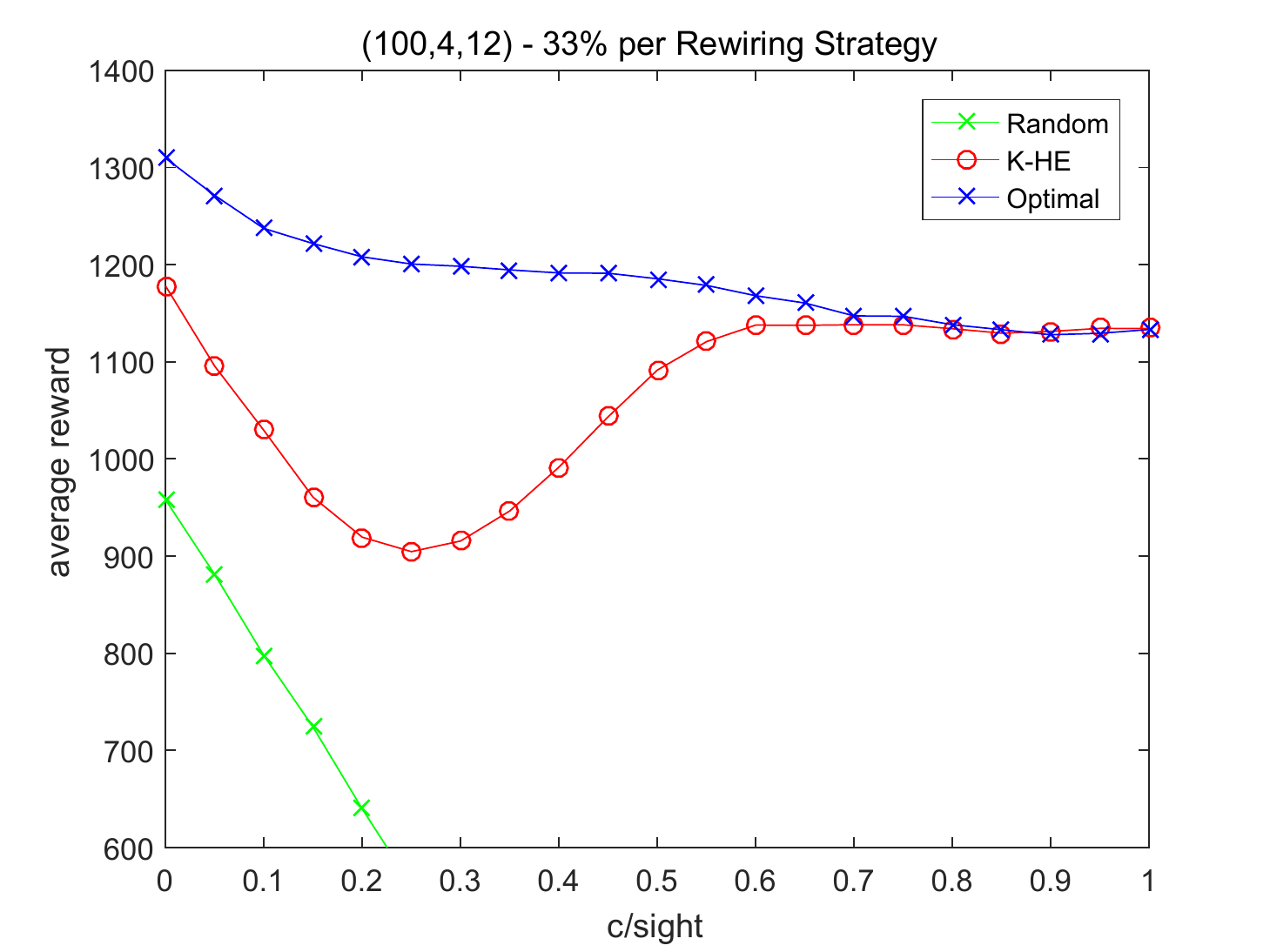}}
\hspace{-0.60cm}
\subfigure[with varying proportions]{
\label{sub4}
\includegraphics[width=0.26\textwidth]{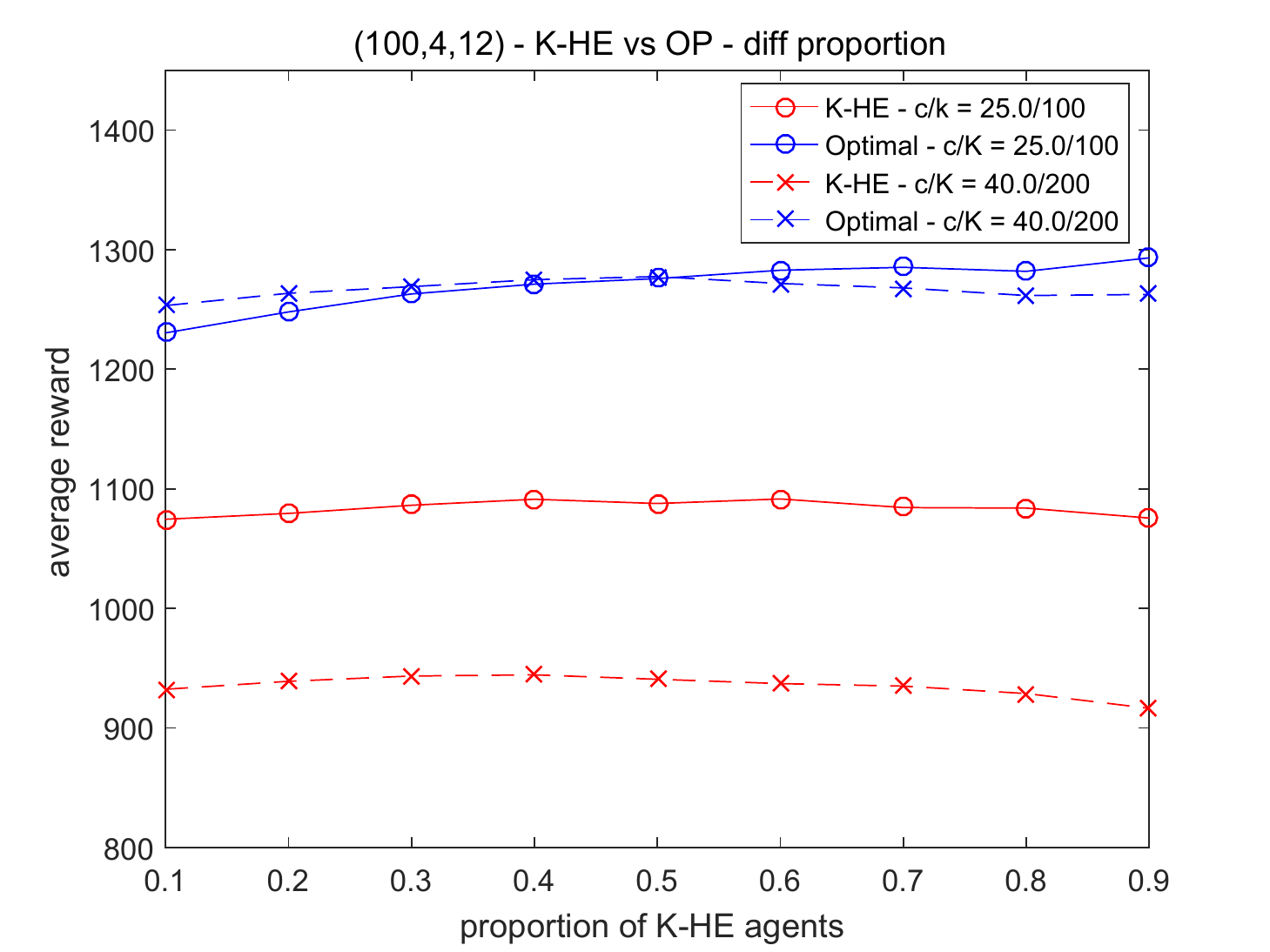}}

\caption{Average payoff of different rewiring strategies generated in 1000-rounds interactions over 100 randomly initialized (100,4,12) networks with $\varphi = 0.01$.}
\label{fig:3}
\end{figure*}

Fig.2(a) shows the average accumulated payoff obtained by each agent with different cost-sight settings.
We can see that the average payoffs for different costs increase rapidly with the growth of sight value and reach the peaks when $c/K$ is within the range of $[0.0,0.2]$.
Intuitively this indicates that an agent with relatively small sight value is short-sighted and may not be willing to rewire any high-risk agent (high cost) to maximize its long-term benefits.
This phenomenon can be explained from Eq.9: the index $\Lambda$ is more likely to be negative with large $c/K$ which means few rewiring decisions can be made with our optimal strategy.
After the peaks, it starts to descend with different speed when the value of $K$ further increases.
This phenomenon indicates that extremely large sight values make optimal estimations become unreliable.
This can also be explained from Eq.9: the small value of $c/K$ leads to the increase of $\Lambda$, and thus increasing the set of candidate agents for rewiring.
In other words, large $K$ is deemed to be over-optimistic and aggressive as any connection may be replaced in the future.
In the extreme case where $K \rightarrow \infty$,
the infinitely long sight makes the rewiring cost meaningless as $c/K \rightarrow 0$.

Based on the above analysis, we can see that given any cost $c$, we should select a reasonable $K$ as the optimal rewiring sight value.
In following experiments, we choose a suitable value of $K$ to let the value of $c/K$ be within the optimal range of $[0.0,0.2]$ for different rewiring costs.

\subsection{Performance Comparison of Rewiring Strategies}
To evaluate the performance of our optimal approach,
we compare our rewiring strategy with two benchmark strategies as follows:
\begin{itemize}
\item $\emph{Random (Ran)}$: Random rewiring is included as the baseline strategy, in which an agent establishes new connection with a potential peer randomly selected each time.
\item \emph{K-sight Highest Expect (K-HE)}:
K-HE is a rational benchmark strategy that rewires the agent
with the highest K-round expected value minus the cost, i.e. $\max_{j \in \bar O} E(x_i^j) - c_i^j/K$.%\footnote{It
%is a K-step variant of Highest Expect (HE) strategy
%which always rewires with a potential peer with the highest one-shot expected payoff.}
\end{itemize}
To the best of our knowledge, there exist works on rewiring in social networks,
but they use rewiring as an additional mechanism instead of designing optimal rewiring strategies
to facilitate multiagent coordination under the social learning framework.
Thus, most of them are are vanilla and cannot be directly applied in our context.

We first evaluate the performance of each rewiring strategy
in different interaction environments with varying parameters $(x,y,z)$ of networks.
%We use $x$, $y$, $z$ separately to denote the size of agents, neighborhood and reachable agents.
Both K-HE and our optimal strategy use the same value of $K$ ($K = 400$) as the rewiring sight value.
We simulate 1000-rounds interactions in pure environments where the population uses the same rewiring strategy.
The average accumulated payoffs obtained by agents using different rewiring strategies are shown in Table 1.

\begin{table}
  \caption{The performance of optimal rewiring strategy and benchmarks with different topologies $c = 20.0$ and $\varphi = 0.01$}
  \label{tab:1}
  \begin{tabular}{cccccl}
    \toprule
    \textbf{No.} & \textbf{(x, y, z)} & \textbf{Rew\_Stg} & \textbf{Avg.} & \textbf{Max.} & \textbf{Min.} \\
    \midrule

	$1$ & $(100,4,8)$ & $Random$ & $978.23$ & $1900.74$ & $355.70$ \\

	$2$ & $(100,4,8)$ & $K-HE$ & $1146.60$ & $2197.96$ & $358.33$ \\

	$3$ & $(100,4,8)$ & $Optimal$ & $1342.78$ & $2261.85$ & $1342.78$ \\
	\rowcolor{mygray}
	$4$ & $(100,4,12)$ & $Random$ & $763.93$ & $1383.97$ & $283.87$ \\
	\rowcolor{mygray}
	$5$ & $(100,4,12)$ & $K-HE$ & $1003.46$ & $1979.67$ & $306.71$ \\
	\rowcolor{mygray}
	$6$ & $(100,4,12)$ & $Optimal$ & $1372.66$ & $2437.97$ & $665.78$ \\

	$7$ & $(100,4,16)$ & $Random$ & $702.26$ & $1290.33$ & $252.51$ \\

	$8$ & $(100,4,16)$ & $K-HE$ & $991.24$ & $1912.27$ & $324.64$ \\

	$9$ & $(100,4,16)$ & $Optimal$ & $1377.95$ & $2516.00$ & $650.10$ \\
	\rowcolor{mygray}
	$10$ & $(500,4,16)$ & $Random$ & $694.34$ & $1434.54$ & $169.40$ \\
	\rowcolor{mygray}
	$11$ & $(500,4,16)$ & $K-HE$ & $993.56$ & $2173.42$ & $211.77$ \\
	\rowcolor{mygray}
	$12$ & $(500,4,16)$ & $Optimal$ & $1373.60$ & $2735.48$ & $484.43$ \\

	$13$ & $(500,8,16)$ & $Random$ & $738.88$ & $1307.37$ & $252.85$ \\

	$14$ & $(500,8,16)$ & $K-HE$ & $1016.85$ & $1768.48$ & $360.16$ \\

	$15$ & $(500,8,16)$ & $Optimal$ & $1177.35$ & $1789.12$ & $593.17$ \\
	\rowcolor{mygray}
	$16$ & $(1000,8,16)$ & $Random$ & $740.29$ & $1375.69$ & $217.80$ \\
	\rowcolor{mygray}
	$17$ & $(1000,8,16)$ & $K-HE$ & $1018.53$ & $1801.04$ & $314.75$ \\
	\rowcolor{mygray}
	$18$ & $(1000,8,16)$ & $Optimal$ & $1170.46$ & $1810.58$ & $567.38$ \\
  \bottomrule
\end{tabular}
\end{table}

First we can observe that our optimal strategy outperforms benchmark strategies in terms of
average, best and worst cases across all different parameter settings,
especially when the size $y$ of neighborhood is relatively small (each neighbor matters).
Second, another observation we can find in No.1 to 9, is that as the size $z$ of reachable agents becomes larger,
the performance of the two benchmark strategies decreases,
while the optimal strategy still shows good and robust performance (actually a slight ascending trend)
against the variation of interaction environments.
This is because our optimal strategy is more likely to pick off better peers when our agents have more rewiring alternatives,
which ensures robust and even better performance.
In contrast, random strategy is more likely to make bad choices.
For K-HE strategy, it only focuses on the highest expected interaction payoff of potential peers
regardless of the network dynamics caused by rewiring.
Hence the loss of optimality becomes even worse when facing more choices.

Another observation is that, comparing the results from No.10 to 15,
we can see that the average payoff for our optimal strategy decreases and the performance gap
between optimal strategy and others becomes smaller as the size $y$ of neighborhood becomes larger.
The is because in our experiments, each agent interacts with a randomly selected neighbor.
Thus, even though our approach always makes the optimal rewiring decision,
the expected payoff during the course of interaction will approach the expectation value of random interactions with the neighborhood,
as the neighborhood size $y$ approaches infinity.
This can be analyzed formally as follows.
For any agent $i$, let $\tilde{v_i}(O_i)$ denotes the agent $i$'s expected value
of random interactions with neighborhood $O_i$ and naturally we have
$\tilde{v_i}(O_i) = 1/|O_i| \sum_{j \in O_i} v_i^j$.
We can see in Eq.1 $v_i^j = \max_{m \in A_j} u_m$ when $p_i^j = 1$, which means agent $i$ and $j$ coordinate on action $m$.
Thus, in the uniform distribution case, where $u_m \sim U(a,b)$ and $a,b \sim U(0,1)$, we easily have
$\lim_{|O_i| \to \infty}E(\tilde{v_i}(O_i)) \rightarrow E(U(0,1))$.

Furthermore,
we compare the performance of our optimal strategy and existing benchmark strategies in different rewiring cost settings.
The value of sight is set to $K = 200$ for both K-HE and the optimal strategy.
We vary rewiring cost value of $c$ in the range of $[0.0,200.0]$.
Fig.2(b) shows the average accumulated payoffs for each rewiring strategy in a pure environment.
We can see that our optimal rewiring strategy outperforms others and acquires higher payoff across almost all $c/K$ value ($[0.0, 0.8]$).
Next we evaluate our optimal rewiring strategy in a competitive environment where each agent is randomly assigned a rewiring strategy (from Random, K-HE and optimal strategy) with equal probabilities and the results are shown in Fig.2(c).
Fig.2(b) and Fig.2(c) are similar and we can see that
our optimal rewiring strategy significantly outperforms existing benchmark strategies for different $c/K$ settings in competitive environments.
As expected, the optimal strategy's payoff starts at the peak when the rewiring cost is zero, and decreases gradually as the cost increases.
The obtained payoff slowly declines until $c/K \approx 0.7$, where the rewiring costs are too high and not rewiring is the optimal choice.

\begin{figure}
\centering

\subfigure[FP vs. JAL vs. JA-WoLF]{
\label{sub1}
\includegraphics[width=0.24\textwidth]{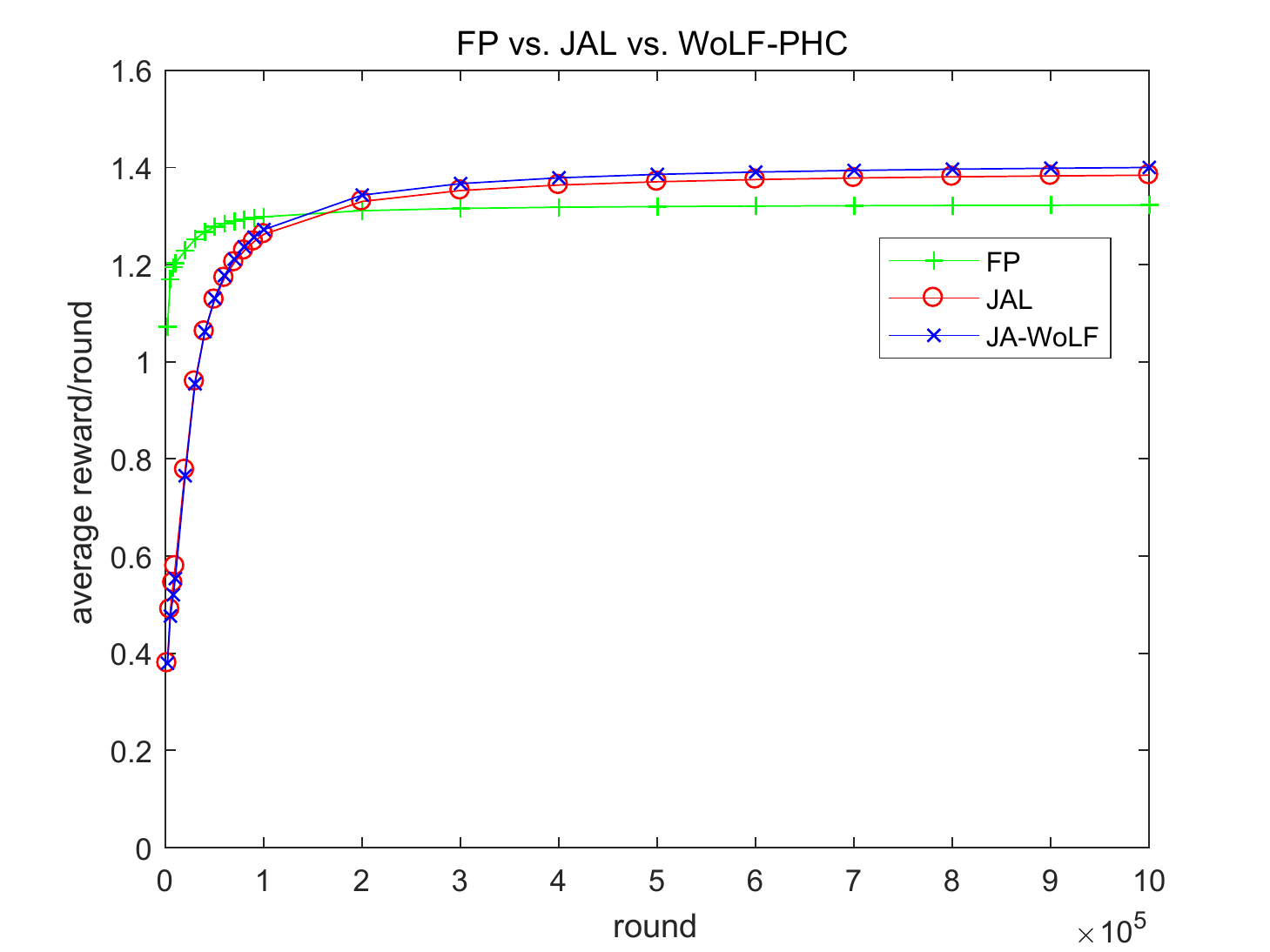}}
\hspace{-0.60cm}
\subfigure[outcomes for coordination]{
\label{sub2}
\includegraphics[width=0.24\textwidth]{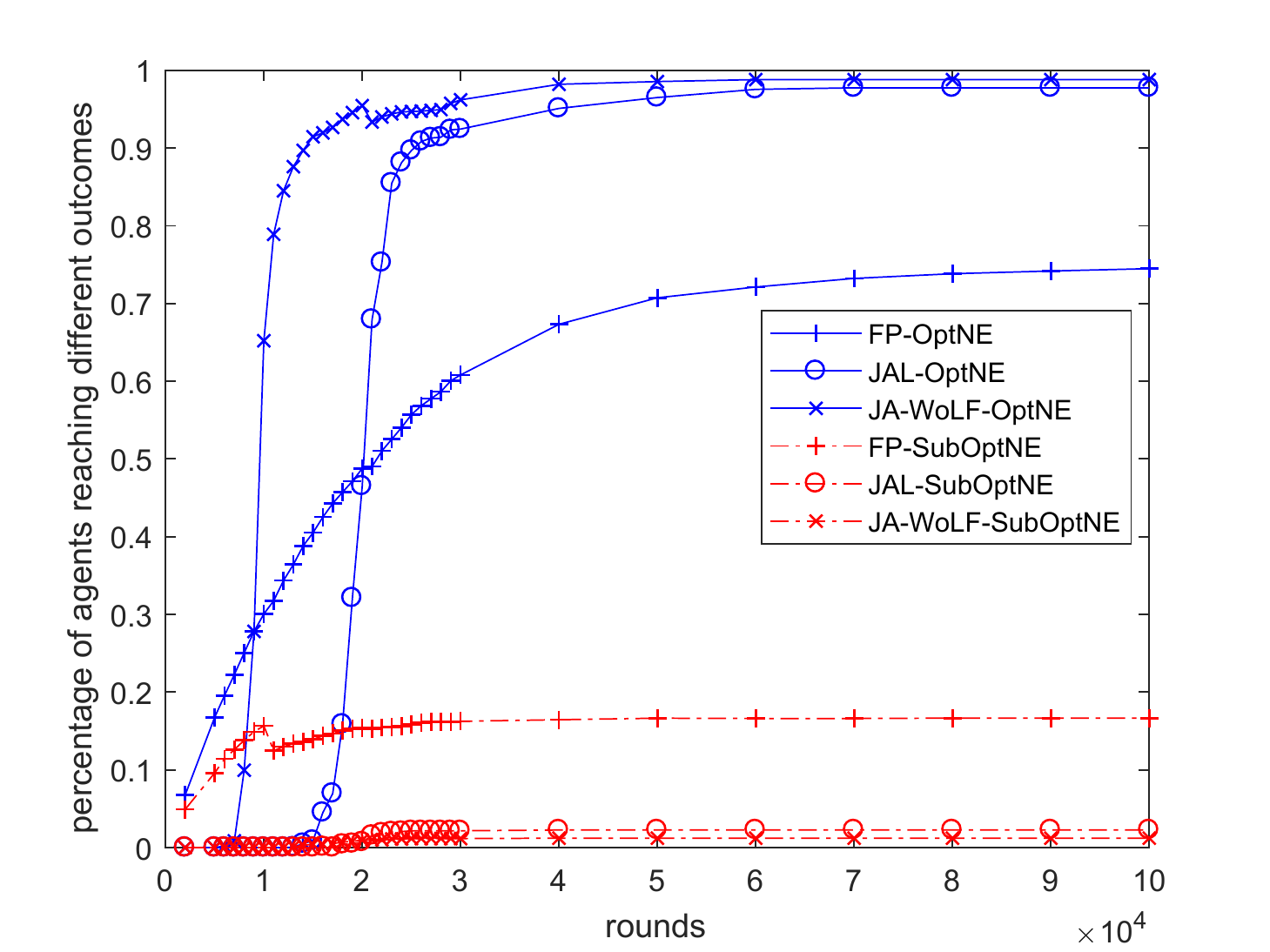}}

\caption{Analysis of learning strategies in (100,4,12) networks with $c = 2000.0$, $\varphi = 0.0001$, $K = 10000.0$ for optimal rewiring strategy in pure environments.}
\label{fig:4}
\end{figure}

The K-HE strategy ranks the second.
It starts with average accumulated payoff near $1200$ and drops significantly to the minimum when $c/K \approx 0.25$.
The reason why K-HE has a good start is the initial rewiring cost is negligible even though its rewiring decision is sub-optimal.
As the rewiring cost increases, the performance gap between K-HE and our optimal strategy becomes even larger.
This is because when $c/K$ is small, there are many rewiring alternatives for both K-HE and our optimal strategy.
For K-HE, it is more likely to pick the sub-optimal one while our optimal strategy always chooses the optimal one.
Further, K-HE's performance gradually improves when $c/K$ further increases and reach the same level as our optimal strategy when $c/K \ge 0.7$.
This is because few alternatives exist for selection as the value of $c/K$ becomes larger and
it is more likely for K-HE to also select the optimal rewiring action by chance.
Therefore, for K-HE strategy,
rewiring with beneficial agents is a safer choice for higher rewiring costs because such agents are sure to give higher payoff;
in contrast, rewiring decisions are not reliable when $c/K$ is within the range of $[0.1,0.5]$.
Overall, the performance difference between K-HE and our optimal strategy is significant, which stems from the fact that
our optimal policy takes future interactions into account when exploring the different potential partners.

The random rewiring method performs the worst since it always rewires a randomly selected peers.
It quickly degenerates and moves off the chart for higher costs because of the rapid increase of the total rewiring costs.
In addition, we vary the proportions of agents using K-HE strategy from $10\%$ to $90\%$ for more competitive environments consisting of agents using only K-HE or optimal strategies, and the results are shown in Fig.2(d).
The random strategy is not considered due to its poor performance.
We can see our optimal strategy significantly prevails K-HE  over all proportion values against $c/K$ settings, which indicates our optimal strategy is robust and can always achieve higher accumulated payoff.

\subsection{Performance Analysis of Different Learning Strategies}
We compare the performance of three representative learning strategies: FP, JAL and JA-WoLF.
Fig.3(a) shows the dynamics of the average single-round payoff for different learning strategies.
It shows the agents using FP strategy can fast reach a good payoff level 
within 10000 rounds and then stay at near 1.3 after 100000 rounds.
JAL and JA-WoLF start with worse performance but outperform FP from near 200000$th$ round
and maintain a 0.1 lead at the value of average per-round reward.
For further explanation, Fig.3(b) shows the dynamics of the percentage of agents reaching optimal Nash equilibrium (OptNE) and sub-optimal equilibrium (SubOptNE) for above three strategies.
We can find that JA-WoLF is the quickest to reach almost 100\% OptNE ($\approx$40000 rounds), followed by JAL ($\approx$60000 rounds).
The population of agents using FP shows a quick start but leads to only over 70\% of OptNE (the rest of 17\% for SubOptNE) which results in its poor behavior during long-run interactions in Fig.3(a).
We hypothesize that FP's bad performance is because its convergence strategy highly depends on the initial strategies of players.
For example, if two players are assigned a high probability of choosing the action which coordinates to the sub-optimal NE, each of them are going to choose the sub-optimal action as a best response which will be reinforced gradually and eventually converge to the sub-optimal NE.
In contrast, JAL and JA-WoLF can always reach almost perfect coordination on the optimal NE.

%%%%%%%%%%%%%%%%%%%%%%%%%%%%%%%%%%%%%%%%%%%%%%%%%%%%%%%%%%%%%%%%%%%%%%%%%%%%%%%%%%%%%%%%%%%%%%%%%%%%%%%%%
%% Conclusion

\section{Conclusion and Future Work}
In this paper, we deal with multiagent coordination problem
in cooperative environments under the social learning framework
with limited rewiring chances available.
We proposed a new rewiring strategy which is optimal and efficient for the agents to maximize accumulated payoff during long-run interactions.
Our empirical results show that our method outperforms other benchmark strategies
in both competitive and pure environments.
Moreover, our method is robust under a
variety of circumstances.
Besides, we analyzed the comparative performance of three representative learning strategies
(i.e., FP, JAL and JA-WoLF) under our social learning framework with optimal rewiring.
The empirical results show that JAL and JA-WoLF outperforms FP on both accumulated payoff 
and the percentage of times agents reaching the optimal Nash equilibria (NE).
In this paper,
we only consider the rewiring problem in cooperative MASs. One worthwhile direction is to investigate how this rewiring strategy can be extended and applied in non-cooperative MASs.
Another research direction is to explicitly consider
how to better utilize the property of the underlying network topology to further facilitate coordination.

%%%%%%%%%%%%%%%%%%%%%%%%%%%%%%%%%%%%%%%%%%%%%%%%%%%%%%%%%%%%%%%%%%%%%%%%%%%%%%%%%%%%%%%%%%%%%%%%%%%%%%%%%
%% bibliography: see CFP for number of permitted pages

\bibliographystyle{ACM-Reference-Format}  % do not change this line!
\bibliography{aaai18}  % put name of your .bib file here


\begin{thebibliography}{00}

%%% the Web address.

\ifx \showCODEN    \undefined \def \showCODEN     #1{\unskip}     \fi
\ifx \showDOI      \undefined \def \showDOI       #1{#1}\fi
\ifx \showISBNx    \undefined \def \showISBNx     #1{\unskip}     \fi
\ifx \showISBNxiii \undefined \def \showISBNxiii  #1{\unskip}     \fi
\ifx \showISSN     \undefined \def \showISSN      #1{\unskip}     \fi
\ifx \showLCCN     \undefined \def \showLCCN      #1{\unskip}     \fi
\ifx \shownote     \undefined \def \shownote      #1{#1}          \fi
\ifx \showarticletitle \undefined \def \showarticletitle #1{#1}   \fi
\ifx \showURL      \undefined \def \showURL       {\relax}        \fi
% invisible to TeX
\providecommand\bibfield[2]{#2}
\providecommand\bibinfo[2]{#2}
\providecommand\natexlab[1]{#1}
\providecommand\showeprint[2][]{arXiv:#2}

\bibitem[\protect\citeauthoryear{Airiau, Sen, and Villatoro}{Airiau
  et~al\mbox{.}}{2014}]%
        {airiau2014emergence}
\bibfield{author}{\bibinfo{person}{St{\'e}phane Airiau},
  \bibinfo{person}{Sandip Sen}, {and} \bibinfo{person}{Daniel Villatoro}.}
  \bibinfo{year}{2014}\natexlab{}.
\newblock \showarticletitle{Emergence of conventions through social learning}.
\newblock \bibinfo{journal}{{\em Autonomous Agents and Multi-Agent Systems\/}}
  \bibinfo{volume}{28}, \bibinfo{number}{5} (\bibinfo{year}{2014}),
  \bibinfo{pages}{779--804}.
\newblock


\bibitem[\protect\citeauthoryear{Albert and Barab{\'a}si}{Albert and
  Barab{\'a}si}{2002}]%
        {albert2002statistical}
\bibfield{author}{\bibinfo{person}{R{\'e}ka Albert} {and}
  \bibinfo{person}{Albert-L{\'a}szl{\'o} Barab{\'a}si}.}
  \bibinfo{year}{2002}\natexlab{}.
\newblock \showarticletitle{Statistical mechanics of complex networks}.
\newblock \bibinfo{journal}{{\em Reviews of modern physics\/}}
  \bibinfo{volume}{74}, \bibinfo{number}{1} (\bibinfo{year}{2002}),
  \bibinfo{pages}{47}.
\newblock


\bibitem[\protect\citeauthoryear{Baarslag, Alan, Gomer, Alam, Perera, Gerding,
  and Schraefel}{Baarslag et~al\mbox{.}}{2017}]%
        {baarslag2017permission}
\bibfield{author}{\bibinfo{person}{Tim Baarslag}, \bibinfo{person}{Alper~T.
  Alan}, \bibinfo{person}{Richard Gomer}, \bibinfo{person}{Mudasser Alam},
  \bibinfo{person}{Charith Perera}, \bibinfo{person}{Enrico~H. Gerding}, {and}
  \bibinfo{person}{M.C. Schraefel}.} \bibinfo{year}{2017}\natexlab{}.
\newblock \showarticletitle{An Automated Negotiation Agent for Permission
  Management}. In \bibinfo{booktitle}{{\em Proceedings of the 2017
  International Conference on Autonomous Agents and Multi-agent Systems}} {\em
  (\bibinfo{series}{AAMAS '17})}. \bibinfo{publisher}{International Foundation
  for Autonomous Agents and Multiagent Systems}, \bibinfo{address}{Richland,
  SC}, \bibinfo{pages}{380--390}.
\newblock
\showURL{%
\url{http://dl.acm.org/citation.cfm?id=3091125.3091184}}


\bibitem[\protect\citeauthoryear{Baarslag and Gerding}{Baarslag and
  Gerding}{2015}]%
        {baarslag2015optimal}
\bibfield{author}{\bibinfo{person}{Tim Baarslag} {and}
  \bibinfo{person}{Enrico~H Gerding}.} \bibinfo{year}{2015}\natexlab{}.
\newblock \showarticletitle{Optimal Incremental Preference Elicitation during
  Negotiation.}. In \bibinfo{booktitle}{{\em IJCAI}}. \bibinfo{pages}{3--9}.
\newblock


\bibitem[\protect\citeauthoryear{Bowling and Veloso}{Bowling and
  Veloso}{2001}]%
        {bowling2001rational}
\bibfield{author}{\bibinfo{person}{Michael Bowling} {and}
  \bibinfo{person}{Manuela Veloso}.} \bibinfo{year}{2001}\natexlab{}.
\newblock \showarticletitle{Rational and convergent learning in stochastic
  games}. In \bibinfo{booktitle}{{\em International joint conference on
  artificial intelligence}}, Vol.~\bibinfo{volume}{17}. LAWRENCE ERLBAUM
  ASSOCIATES LTD, \bibinfo{pages}{1021--1026}.
\newblock


\bibitem[\protect\citeauthoryear{Claus and Boutilier}{Claus and
  Boutilier}{1998}]%
        {claus1998dynamics}
\bibfield{author}{\bibinfo{person}{Caroline Claus} {and} \bibinfo{person}{Craig
  Boutilier}.} \bibinfo{year}{1998}\natexlab{}.
\newblock \showarticletitle{The dynamics of reinforcement learning in
  cooperative multiagent systems}.
\newblock \bibinfo{journal}{{\em AAAI/IAAI\/}}  \bibinfo{volume}{1998}
  (\bibinfo{year}{1998}), \bibinfo{pages}{746--752}.
\newblock


\bibitem[\protect\citeauthoryear{Crow, Widjaja, Kim, and Sakai}{Crow
  et~al\mbox{.}}{1997}]%
        {crow1997ieee}
\bibfield{author}{\bibinfo{person}{Brian~P Crow}, \bibinfo{person}{Indra
  Widjaja}, \bibinfo{person}{Jeong~Geun Kim}, {and} \bibinfo{person}{Prescott~T
  Sakai}.} \bibinfo{year}{1997}\natexlab{}.
\newblock \showarticletitle{IEEE 802.11 wireless local area networks}.
\newblock \bibinfo{journal}{{\em IEEE Communications magazine\/}}
  \bibinfo{volume}{35}, \bibinfo{number}{9} (\bibinfo{year}{1997}),
  \bibinfo{pages}{116--126}.
\newblock


\bibitem[\protect\citeauthoryear{Damer and Gini}{Damer and Gini}{2008}]%
        {damer2008achieving}
\bibfield{author}{\bibinfo{person}{Steven Damer} {and} \bibinfo{person}{Maria~L
  Gini}.} \bibinfo{year}{2008}\natexlab{}.
\newblock \showarticletitle{Achieving Cooperation in a Minimally Constrained
  Environment.}. In \bibinfo{booktitle}{{\em AAAI}}. \bibinfo{pages}{57--62}.
\newblock


\bibitem[\protect\citeauthoryear{Dekker}{Dekker}{2007}]%
        {dekker2007realistic}
\bibfield{author}{\bibinfo{person}{AH Dekker}.}
  \bibinfo{year}{2007}\natexlab{}.
\newblock \showarticletitle{Realistic social networks for simulation using
  network rewiring}. In \bibinfo{booktitle}{{\em International Congress on
  Modelling and Simulation}}. \bibinfo{pages}{677--683}.
\newblock


\bibitem[\protect\citeauthoryear{Ellison et~al\mbox{.}}{Ellison
  et~al\mbox{.}}{2007a}]%
        {ellison2007social}
\bibfield{author}{\bibinfo{person}{Nicole~B Ellison} {et~al\mbox{.}}}
  \bibinfo{year}{2007}\natexlab{a}.
\newblock \showarticletitle{Social network sites: Definition, history, and
  scholarship}.
\newblock \bibinfo{journal}{{\em Journal of Computer-Mediated Communication\/}}
  \bibinfo{volume}{13}, \bibinfo{number}{1} (\bibinfo{year}{2007}),
  \bibinfo{pages}{210--230}.
\newblock


\bibitem[\protect\citeauthoryear{Ellison, Steinfield, and Lampe}{Ellison
  et~al\mbox{.}}{2007b}]%
        {ellison2007benefits}
\bibfield{author}{\bibinfo{person}{Nicole~B Ellison}, \bibinfo{person}{Charles
  Steinfield}, {and} \bibinfo{person}{Cliff Lampe}.}
  \bibinfo{year}{2007}\natexlab{b}.
\newblock \showarticletitle{The benefits of Facebook “friends:” Social
  capital and college students’ use of online social network sites}.
\newblock \bibinfo{journal}{{\em Journal of Computer-Mediated Communication\/}}
  \bibinfo{volume}{12}, \bibinfo{number}{4} (\bibinfo{year}{2007}),
  \bibinfo{pages}{1143--1168}.
\newblock


\bibitem[\protect\citeauthoryear{Gerding, Larson, Rogers, and Jennings}{Gerding
  et~al\mbox{.}}{2009}]%
        {gerding2009mechanism}
\bibfield{author}{\bibinfo{person}{Enrico~H Gerding}, \bibinfo{person}{Kate
  Larson}, \bibinfo{person}{Alex Rogers}, {and} \bibinfo{person}{Nicholas~R
  Jennings}.} \bibinfo{year}{2009}\natexlab{}.
\newblock \showarticletitle{Mechanism design for task procurement with flexible
  quality of service}. In \bibinfo{booktitle}{{\em International Workshop on
  Service-Oriented Computing: Agents, Semantics, and Engineering}}. Springer,
  \bibinfo{pages}{12--23}.
\newblock


\bibitem[\protect\citeauthoryear{Griffiths and Luck}{Griffiths and
  Luck}{2010}]%
        {griffiths2010changing}
\bibfield{author}{\bibinfo{person}{Nathan Griffiths} {and}
  \bibinfo{person}{Michael Luck}.} \bibinfo{year}{2010}\natexlab{}.
\newblock \showarticletitle{Changing neighbours: improving tag-based
  cooperation}. In \bibinfo{booktitle}{{\em Proceedings of the 9th
  International Conference on Autonomous Agents and Multiagent Systems: volume
  1-Volume 1}}. International Foundation for Autonomous Agents and Multiagent
  Systems, \bibinfo{pages}{249--256}.
\newblock


\bibitem[\protect\citeauthoryear{Hales and Edmonds}{Hales and Edmonds}{2005}]%
        {hales2005applying}
\bibfield{author}{\bibinfo{person}{David Hales} {and} \bibinfo{person}{Bruce
  Edmonds}.} \bibinfo{year}{2005}\natexlab{}.
\newblock \showarticletitle{Applying a socially inspired technique (tags) to
  improve cooperation in P2P networks}.
\newblock \bibinfo{journal}{{\em IEEE Transactions on Systems, Man, and
  Cybernetics-Part A: Systems and Humans\/}} \bibinfo{volume}{35},
  \bibinfo{number}{3} (\bibinfo{year}{2005}), \bibinfo{pages}{385--395}.
\newblock


\bibitem[\protect\citeauthoryear{Hao, Huang, Cai, and Leung}{Hao
  et~al\mbox{.}}{2014}]%
        {hao2014reinforcement}
\bibfield{author}{\bibinfo{person}{Jianye Hao}, \bibinfo{person}{Dongping
  Huang}, \bibinfo{person}{Yi Cai}, {and} \bibinfo{person}{Ho-Fung Leung}.}
  \bibinfo{year}{2014}\natexlab{}.
\newblock \showarticletitle{Reinforcement social learning of coordination in
  networked cooperative multiagent systems}. In \bibinfo{booktitle}{{\em AAAI
  workshop on multiagent interaction without prior coordination (MIPC 2014)}}.
\newblock


\bibitem[\protect\citeauthoryear{Hao, Huang, Cai, and Leung}{Hao
  et~al\mbox{.}}{2017}]%
        {hao2017dynamics}
\bibfield{author}{\bibinfo{person}{Jianye Hao}, \bibinfo{person}{Dongping
  Huang}, \bibinfo{person}{Yi Cai}, {and} \bibinfo{person}{Ho-fung Leung}.}
  \bibinfo{year}{2017}\natexlab{}.
\newblock \showarticletitle{The dynamics of reinforcement social learning in
  networked cooperative multiagent systems}.
\newblock \bibinfo{journal}{{\em Engineering Applications of Artificial
  Intelligence\/}}  \bibinfo{volume}{58} (\bibinfo{year}{2017}),
  \bibinfo{pages}{111--122}.
\newblock


\bibitem[\protect\citeauthoryear{Hao and Leung}{Hao and Leung}{2013}]%
        {hao2013dynamics}
\bibfield{author}{\bibinfo{person}{Jianye Hao} {and} \bibinfo{person}{Ho-fung
  Leung}.} \bibinfo{year}{2013}\natexlab{}.
\newblock \showarticletitle{The Dynamics of Reinforcement Social Learning in
  Cooperative Multiagent Systems.}. In \bibinfo{booktitle}{{\em IJCAI}},
  Vol.~\bibinfo{volume}{13}. \bibinfo{pages}{184--190}.
\newblock


\bibitem[\protect\citeauthoryear{Hao, Leung, and Ming}{Hao
  et~al\mbox{.}}{2015}]%
        {hao2015multiagent}
\bibfield{author}{\bibinfo{person}{Jianye Hao}, \bibinfo{person}{Ho-Fung
  Leung}, {and} \bibinfo{person}{Zhong Ming}.} \bibinfo{year}{2015}\natexlab{}.
\newblock \showarticletitle{Multiagent reinforcement social learning toward
  coordination in cooperative multiagent systems}.
\newblock \bibinfo{journal}{{\em ACM Transactions on Autonomous and Adaptive
  Systems (TAAS)\/}} \bibinfo{volume}{9}, \bibinfo{number}{4}
  (\bibinfo{year}{2015}), \bibinfo{pages}{20}.
\newblock


\bibitem[\protect\citeauthoryear{Kapetanakis and Kudenko}{Kapetanakis and
  Kudenko}{2002}]%
        {kapetanakis2002reinforcement}
\bibfield{author}{\bibinfo{person}{Spiros Kapetanakis} {and}
  \bibinfo{person}{Daniel Kudenko}.} \bibinfo{year}{2002}\natexlab{}.
\newblock \showarticletitle{Reinforcement learning of coordination in
  cooperative multi-agent systems}.
\newblock \bibinfo{journal}{{\em AAAI/IAAI\/}}  \bibinfo{volume}{2002}
  (\bibinfo{year}{2002}), \bibinfo{pages}{326--331}.
\newblock


\bibitem[\protect\citeauthoryear{Kwak, Lee, Park, and Moon}{Kwak
  et~al\mbox{.}}{2010}]%
        {kwak2010twitter}
\bibfield{author}{\bibinfo{person}{Haewoon Kwak}, \bibinfo{person}{Changhyun
  Lee}, \bibinfo{person}{Hosung Park}, {and} \bibinfo{person}{Sue Moon}.}
  \bibinfo{year}{2010}\natexlab{}.
\newblock \showarticletitle{What is Twitter, a social network or a news
  media?}. In \bibinfo{booktitle}{{\em Proceedings of the 19th international
  conference on World wide web}}. ACM, \bibinfo{pages}{591--600}.
\newblock


\bibitem[\protect\citeauthoryear{Lauer and Riedmiller}{Lauer and
  Riedmiller}{2000}]%
        {lauer2000algorithm}
\bibfield{author}{\bibinfo{person}{Martin Lauer} {and} \bibinfo{person}{Martin
  Riedmiller}.} \bibinfo{year}{2000}\natexlab{}.
\newblock \showarticletitle{An algorithm for distributed reinforcement learning
  in cooperative multi-agent systems}. In \bibinfo{booktitle}{{\em In
  Proceedings of the Seventeenth International Conference on Machine
  Learning}}. Citeseer.
\newblock


\bibitem[\protect\citeauthoryear{Matignon, Laurent, and Le~Fort-Piat}{Matignon
  et~al\mbox{.}}{2008}]%
        {matignon2008study}
\bibfield{author}{\bibinfo{person}{La{\"e}titia Matignon},
  \bibinfo{person}{Guillaume Laurent}, {and} \bibinfo{person}{Nadine
  Le~Fort-Piat}.} \bibinfo{year}{2008}\natexlab{}.
\newblock \showarticletitle{A study of FMQ heuristic in cooperative multi-agent
  games.}. In \bibinfo{booktitle}{{\em The 7th International Conference on
  Autonomous Agents and Multiagent Systems. Workshop 10: Multi-Agent Sequential
  Decision Making in Uncertain Multi-Agent Domains, aamas' 08.}},
  Vol.~\bibinfo{volume}{1}. \bibinfo{pages}{77--91}.
\newblock


\bibitem[\protect\citeauthoryear{Matignon, Laurent, and Le~Fort-Piat}{Matignon
  et~al\mbox{.}}{2012}]%
        {matignon2012independent}
\bibfield{author}{\bibinfo{person}{Laetitia Matignon},
  \bibinfo{person}{Guillaume~J Laurent}, {and} \bibinfo{person}{Nadine
  Le~Fort-Piat}.} \bibinfo{year}{2012}\natexlab{}.
\newblock \showarticletitle{Independent reinforcement learners in cooperative
  Markov games: a survey regarding coordination problems}.
\newblock \bibinfo{journal}{{\em The Knowledge Engineering Review\/}}
  \bibinfo{volume}{27}, \bibinfo{number}{1} (\bibinfo{year}{2012}),
  \bibinfo{pages}{1--31}.
\newblock


\bibitem[\protect\citeauthoryear{Mihaylov, Tuyls, and Now{\'e}}{Mihaylov
  et~al\mbox{.}}{2014}]%
        {mihaylov2014decentralized}
\bibfield{author}{\bibinfo{person}{Mihail Mihaylov}, \bibinfo{person}{Karl
  Tuyls}, {and} \bibinfo{person}{Ann Now{\'e}}.}
  \bibinfo{year}{2014}\natexlab{}.
\newblock \showarticletitle{A decentralized approach for convention emergence
  in multi-agent systems}.
\newblock \bibinfo{journal}{{\em Autonomous Agents and Multi-Agent Systems\/}}
  \bibinfo{volume}{28}, \bibinfo{number}{5} (\bibinfo{year}{2014}),
  \bibinfo{pages}{749--778}.
\newblock


\bibitem[\protect\citeauthoryear{Panait and Luke}{Panait and Luke}{2005}]%
        {panait2005cooperative}
\bibfield{author}{\bibinfo{person}{Liviu Panait} {and} \bibinfo{person}{Sean
  Luke}.} \bibinfo{year}{2005}\natexlab{}.
\newblock \showarticletitle{Cooperative multi-agent learning: The state of the
  art}.
\newblock \bibinfo{journal}{{\em Autonomous agents and multi-agent systems\/}}
  \bibinfo{volume}{11}, \bibinfo{number}{3} (\bibinfo{year}{2005}),
  \bibinfo{pages}{387--434}.
\newblock


\bibitem[\protect\citeauthoryear{Panait, Sullivan, and Luke}{Panait
  et~al\mbox{.}}{2006}]%
        {panait2006lenient}
\bibfield{author}{\bibinfo{person}{Liviu Panait}, \bibinfo{person}{Keith
  Sullivan}, {and} \bibinfo{person}{Sean Luke}.}
  \bibinfo{year}{2006}\natexlab{}.
\newblock \showarticletitle{Lenient learners in cooperative multiagent
  systems}. In \bibinfo{booktitle}{{\em Proceedings of the fifth international
  joint conference on Autonomous agents and multiagent systems}}. ACM,
  \bibinfo{pages}{801--803}.
\newblock


\bibitem[\protect\citeauthoryear{Peleteiro, Burguillo, and Chong}{Peleteiro
  et~al\mbox{.}}{2014}]%
        {peleteiro2014exploring}
\bibfield{author}{\bibinfo{person}{Ana Peleteiro}, \bibinfo{person}{Juan~C
  Burguillo}, {and} \bibinfo{person}{Siang~Yew Chong}.}
  \bibinfo{year}{2014}\natexlab{}.
\newblock \showarticletitle{Exploring indirect reciprocity in complex networks
  using coalitions and rewiring}. In \bibinfo{booktitle}{{\em Proceedings of
  the 2014 international conference on Autonomous agents and multi-agent
  systems}}. International Foundation for Autonomous Agents and Multiagent
  Systems, \bibinfo{pages}{669--676}.
\newblock


\bibitem[\protect\citeauthoryear{Sen and Airiau}{Sen and Airiau}{2007}]%
        {sen2007emergence}
\bibfield{author}{\bibinfo{person}{Sandip Sen} {and}
  \bibinfo{person}{St{\'e}phane Airiau}.} \bibinfo{year}{2007}\natexlab{}.
\newblock \showarticletitle{Emergence of norms through social learning.}. In
  \bibinfo{booktitle}{{\em IJCAI}}, Vol.~\bibinfo{volume}{1507}.
  \bibinfo{pages}{1512}.
\newblock


\bibitem[\protect\citeauthoryear{Villatoro, Sabater-Mir, and Sen}{Villatoro
  et~al\mbox{.}}{2011}]%
        {villatoro2011social}
\bibfield{author}{\bibinfo{person}{Daniel Villatoro}, \bibinfo{person}{Jordi
  Sabater-Mir}, {and} \bibinfo{person}{Sandip Sen}.}
  \bibinfo{year}{2011}\natexlab{}.
\newblock \showarticletitle{Social instruments for robust convention
  emergence}. In \bibinfo{booktitle}{{\em IJCAI}}, Vol.~\bibinfo{volume}{11}.
  \bibinfo{pages}{420--425}.
\newblock


\bibitem[\protect\citeauthoryear{Watts and Strogatz}{Watts and
  Strogatz}{1998}]%
        {watts1998collective}
\bibfield{author}{\bibinfo{person}{Duncan~J Watts} {and}
  \bibinfo{person}{Steven~H Strogatz}.} \bibinfo{year}{1998}\natexlab{}.
\newblock \showarticletitle{Collective dynamics of'small-world'networks}.
\newblock \bibinfo{journal}{{\em nature\/}} \bibinfo{volume}{393},
  \bibinfo{number}{6684} (\bibinfo{year}{1998}), \bibinfo{pages}{440}.
\newblock


\bibitem[\protect\citeauthoryear{Weitzman}{Weitzman}{1979}]%
        {weitzman1979optimal}
\bibfield{author}{\bibinfo{person}{Martin~L Weitzman}.}
  \bibinfo{year}{1979}\natexlab{}.
\newblock \showarticletitle{Optimal search for the best alternative}.
\newblock \bibinfo{journal}{{\em Econometrica: Journal of the Econometric
  Society\/}} (\bibinfo{year}{1979}), \bibinfo{pages}{641--654}.
\newblock


\bibitem[\protect\citeauthoryear{Yu, Zhang, Ren, and Luo}{Yu
  et~al\mbox{.}}{2013}]%
        {yu2013emergence}
\bibfield{author}{\bibinfo{person}{Chao Yu}, \bibinfo{person}{Minjie Zhang},
  \bibinfo{person}{Fenghui Ren}, {and} \bibinfo{person}{Xudong Luo}.}
  \bibinfo{year}{2013}\natexlab{}.
\newblock \showarticletitle{Emergence of social norms through collective
  learning in networked agent societies}. In \bibinfo{booktitle}{{\em
  Proceedings of the 2013 international conference on Autonomous agents and
  multi-agent systems}}. International Foundation for Autonomous Agents and
  Multiagent Systems, \bibinfo{pages}{475--482}.
\newblock


\bibitem[\protect\citeauthoryear{Zhang and Lesser}{Zhang and Lesser}{2013}]%
        {zhang2013coordinating}
\bibfield{author}{\bibinfo{person}{Chongjie Zhang} {and}
  \bibinfo{person}{Victor Lesser}.} \bibinfo{year}{2013}\natexlab{}.
\newblock \showarticletitle{Coordinating multi-agent reinforcement learning
  with limited communication}. In \bibinfo{booktitle}{{\em Proceedings of the
  2013 international conference on Autonomous agents and multi-agent systems}}.
  International Foundation for Autonomous Agents and Multiagent Systems,
  \bibinfo{pages}{1101--1108}.
\newblock


\end{thebibliography}

\end{document}